\pdfoutput=1
\documentclass[a4paper,reprint,twocolumn,notitlepage,groupedaddress,aip,nofootinbib]{revtex4-2}
\usepackage[left=1.5cm,right=1.5cm,top=1cm,bottom=1.5cm,includeheadfoot]{geometry}
\usepackage[english]{babel}
\usepackage[T1]{fontenc}
\usepackage{tgtermes} 
\usepackage{tgheros} 

\usepackage{amssymb}
\usepackage{amsmath}
\usepackage{amsthm}
\usepackage{mathtools}
\usepackage{mathrsfs}
\usepackage{bm}
\usepackage{slashed}
\usepackage[dvipsnames]{xcolor}
\usepackage[shortlabels]{enumitem}
\usepackage[normalem]{ulem}
\PassOptionsToPackage{hyphens}{url} 
\usepackage[breaklinks=true]{hyperref}
\bibpunct{[}{]}{;}{n}{}{} 
\usepackage{graphicx}
\usepackage[font=small,labelfont=bf]{caption}
\usepackage{subcaption}
\usepackage{nicematrix}

\usepackage{tikz}
\usetikzlibrary{
    shapes.geometric,
    calc,
    arrows.meta,
    fit,
    backgrounds,
    positioning,
    patterns,
    decorations.pathreplacing
}
\tikzset{
    graphnode/.style={draw,circle,fill=SkyBlue,draw=black}
}

\theoremstyle{plain}

\theoremstyle{definition}

\theoremstyle{remark}

\newcommand{\R}{\mathbb{R}}
\newcommand{\C}{\mathbb{C}}

\newcommand{\B}{\mathbb{B}}
\newcommand{\rmd}{\mathrm{d}}
\newcommand{\rme}{\mathrm{e}}
\newcommand{\rmi}{\mathrm{i}}
\newcommand{\rms}{\mathrm{s}} 

\renewcommand{\H}{\mathcal{H}}
\newcommand{\BB}{\mathbf{B}}
\newcommand{\mm}{\mathbf{m}}
\newcommand{\xx}{\mathbf{x}}
\newcommand{\ssigma}{\bm{\sigma}}
\newcommand{\up}{\uparrow}
\newcommand{\down}{\downarrow}
\newcommand{\Dens}{\mathcal{D}}
\newcommand{\Pot}{\mathcal{V}}
\newcommand{\DensNonDeg}{\Dens^\mathsf{\neg D}}
\newcommand{\PotNonDeg}{\Pot^\mathsf{\neg D}} 
\newcommand{\DensDeg}{\Dens^\mathsf{D}}
\newcommand{\PotDeg}{\Pot^\mathsf{D}} 
\newcommand{\DensNonUv}{\Dens^\mathsf{m}} 
\newcommand{\rhotarget}{\rho_\odot}
\newcommand\orho[1]{\rho_{\overline{#1}}}
\newcommand\oirho[2]{\rho_{\overline{#1},#2}}
\newcommand{\Hxc}{\textrm{Hxc}}
\newcommand{\s}{\mathrm{s}}

\newcommand{\aop}{\hat{a}}
\newcommand{\aopdag}{\aop^\dagger}
\newcommand{\rhoop}{\hat{\rho}}
\newcommand{\mop}{\hat{\bm{m}}}
\DeclareMathOperator{\Tr}{Tr}

\renewcommand{\Re}{\mathop\mathsf{Re}}
\renewcommand{\Im}{\mathop\mathsf{Im}}

\begin{document}
\title{
    Geometrical Perspective on Spin-Lattice Density-Functional Theory
}

\author{Markus Penz}
\affiliation{Max Planck Institute for the Structure and Dynamics of Matter and Center for Free-Electron Laser Science, Hamburg, Germany}
\affiliation{Department of Computer Science, Oslo Metropolitan University, Oslo, Norway}
\email{m.penz@inter.at}

\author{Robert van Leeuwen}
\affiliation{Department of Physics, Nanoscience Center, University of Jyv\"askyl\"a, Finland}

\begin{abstract}
    A recently developed viewpoint on the fundamentals of density-functional theory for finite interacting spin-lattice systems that centers around the notion of degeneracy regions is presented. It allows for an entirely geometrical description of the Hohenberg--Kohn theorem and $v$-representability. The phenomena receive exemplification by an Anderson impurity model and other small-lattice examples. The case of adiabatic change and the time-dependent setting are examined as well.
\end{abstract}

\maketitle

\vspace{-2em}
{\small
\tableofcontents
}

\section{Introduction}
\label{sec:intro}

Within the field of density-functional theory (DFT), lattice models are frequently considered to obtain rigorous theoretical statements in cases where the continuum formulation proves too difficult~\cite{katriel1981mapping,KohnPRL,englisch2,CCR1985,schonhammer1987discontinuity}, or to treat physical situations in an approximate way~\cite{Gunnarsson1986,schonhammer1995density,xianlong2006,Xianlong2012,Carrascal2012,saubanere2014lattice,carrascal2015hubbard,ijas2010lattice,Coe2019}.
When considering a finite lattice, this consists of $M$ lattice sites that serve as the positional degrees-of-freedom for $N$ fermionic quantum particles. 
Such settings arise directly from a space discretization or are introduced as a model in themselves, often including spin and on-site interaction in the form of a Hubbard model~\cite{Qin2022}.
Since they are interpreted as simplifications of the continuum case, it is commonly assumed that any statement from continuum DFT equally holds in the lattice setting.
This adoption of known results applies especially to the Hohenberg--Kohn (HK) theorem~\cite{Hohenberg-Kohn1964}, which states that a ground-state ($v$-representable) density is produced by a unique (up to a constant) potential. 
Yet \citet{CCR1985}, after their proof that every density on an (even infinite) lattice with $0<\rho_i<1$ can be produced by a well-defined potential ($v$-representability), carefully remarked that: ``The HK theorem for fermions at zero temperature remains an open problem.''

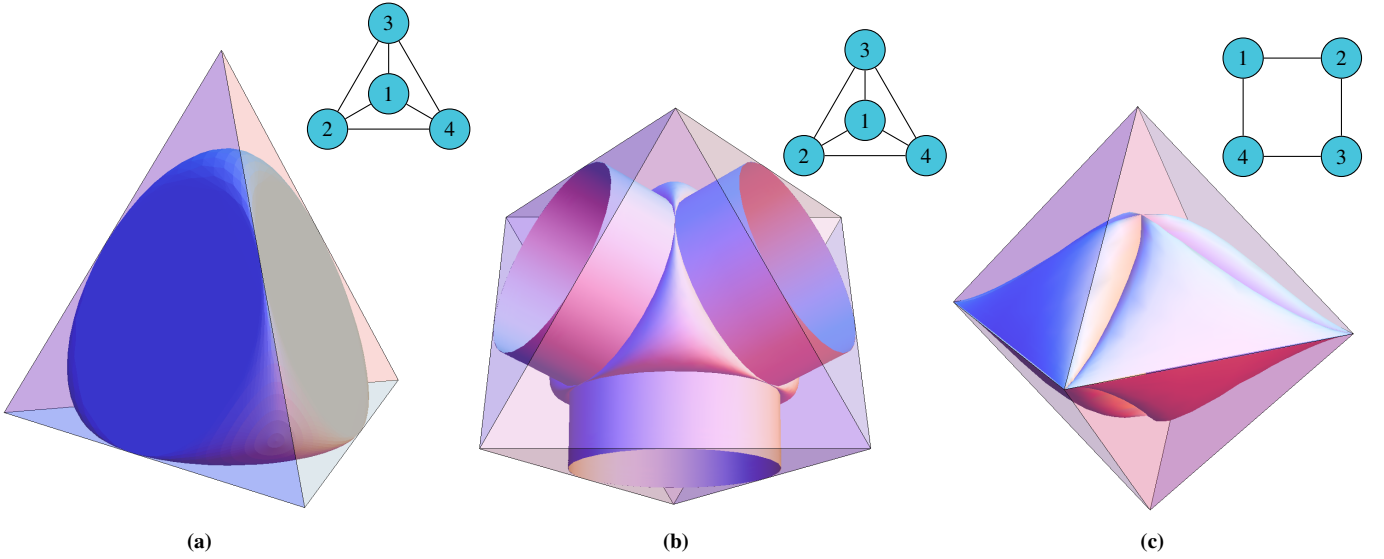
\begin{figure*}[ht]
\begin{subfigure}{.3\textwidth}
    \centering
    \begin{tikzpicture}
        \node[inner sep=0pt] at (0,0)
            {\includegraphics[width=1\columnwidth]{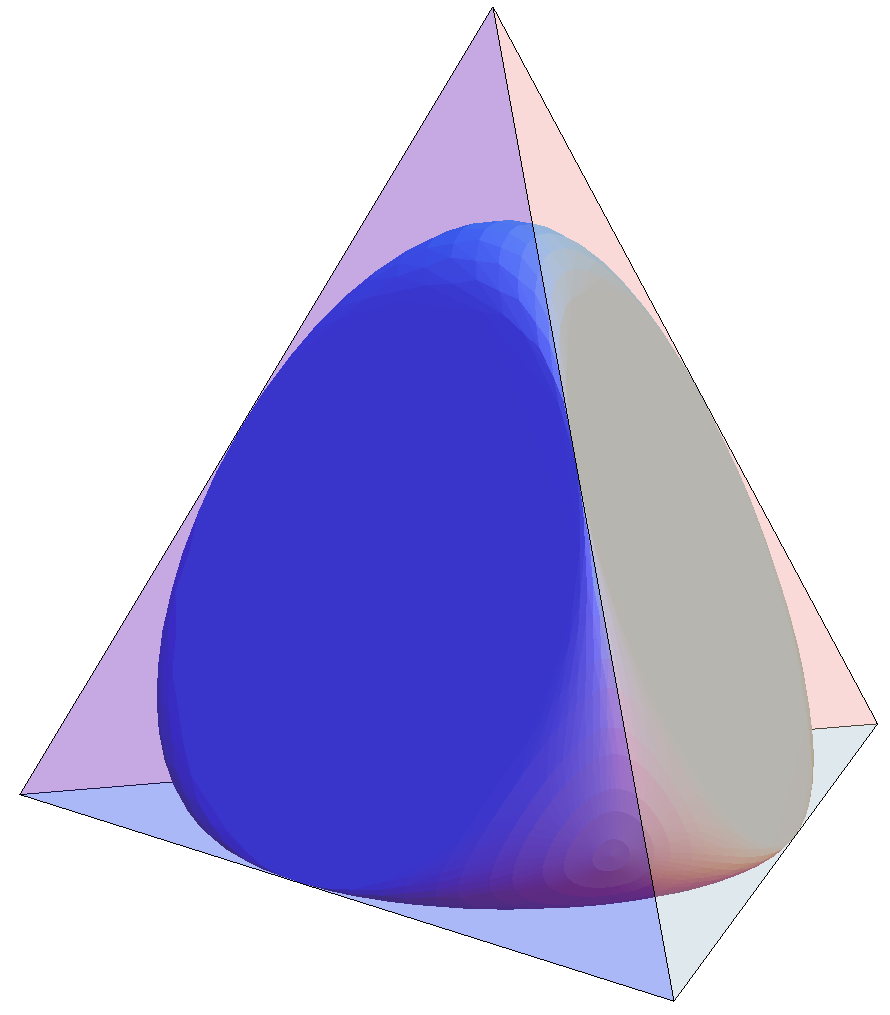}};
        \def\CX{2.5};
        \def\CY{2};
        \def\L{0.7};
        \coordinate (C1) at (0+\CX,{2/3*\L+\CY});
        \coordinate (C2) at ({-2/sqrt(3)*\L+\CX},0+\CY);
        \coordinate (C3) at (0+\CX,{2*\L+\CY});
        \coordinate (C4) at ({2/sqrt(3)*\L+\CX},0+\CY);
        \foreach \m in {1,...,4}
            \node[graphnode,font=\footnotesize] (N\m) at (C\m) {\m};
        \draw (N1) -- (N2);
        \draw (N1) -- (N3);
        \draw (N1) -- (N4);
        \draw (N2) -- (N3);
        \draw (N3) -- (N4);
        \draw (N4) -- (N2);
    \end{tikzpicture}
    \caption{}
\end{subfigure}
\hfill
\begin{subfigure}{.3\textwidth}
    \centering
    \begin{tikzpicture}
        \node[inner sep=0pt] at (0,0)
            {\includegraphics[width=1\columnwidth]{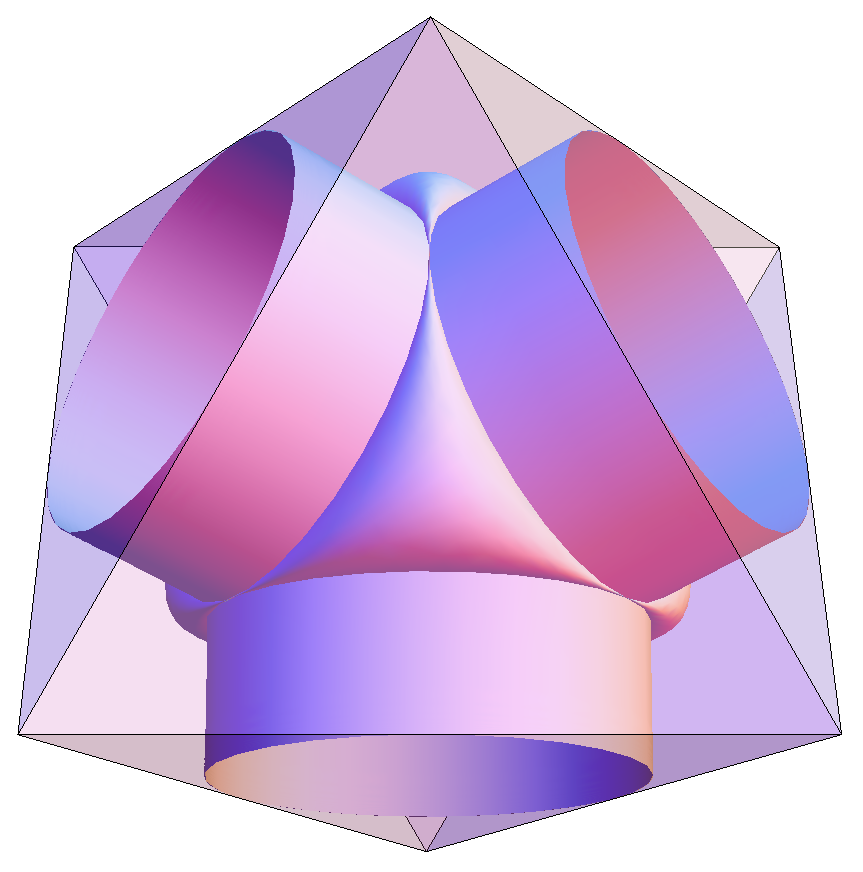}};
        \def\CX{2.5};
        \def\CY{2};
        \def\L{0.7};
        \coordinate (C1) at (0+\CX,{2/3*\L+\CY});
        \coordinate (C2) at ({-2/sqrt(3)*\L+\CX},0+\CY);
        \coordinate (C3) at (0+\CX,{2*\L+\CY});
        \coordinate (C4) at ({2/sqrt(3)*\L+\CX},0+\CY);
        \foreach \m in {1,...,4}
            \node[graphnode,font=\footnotesize] (N\m) at (C\m) {\m};
        \draw (N1) -- (N2);
        \draw (N1) -- (N3);
        \draw (N1) -- (N4);
        \draw (N2) -- (N3);
        \draw (N3) -- (N4);
        \draw (N4) -- (N2);
    \end{tikzpicture}
    \caption{}
\end{subfigure}
\hfill
\begin{subfigure}{.3\textwidth}
    \centering
    \begin{tikzpicture}
        \node[inner sep=0pt] at (0,0)
            {\includegraphics[width=1\columnwidth]{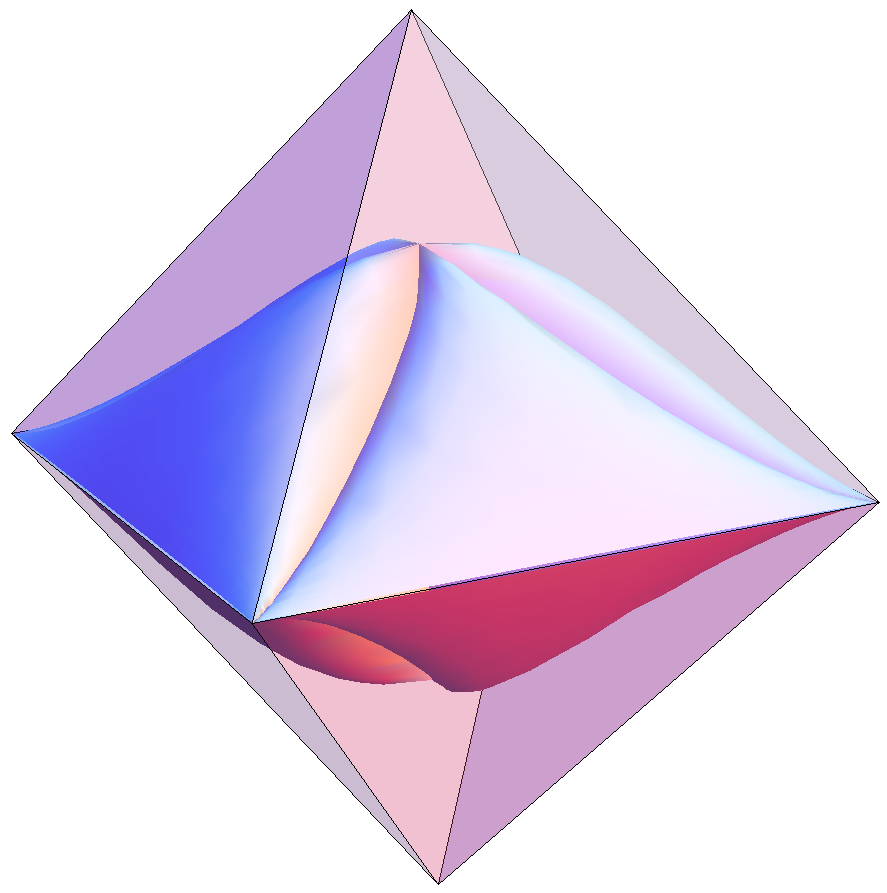}};
        \def\CX{1.2};
        \def\CY{2};
        \def\L{1.3};
        \coordinate (C4) at (0+\CX,0+\CY);
        \coordinate (C3) at (\L+\CX,0+\CY);
        \coordinate (C2) at (\L+\CX,\L+\CY);
        \coordinate (C1) at (0+\CX,\L+\CY);
        \foreach \m in {1,...,4}
            \node[graphnode,font=\footnotesize] (N\m) at (C\m) {\m};
        \draw (N1) -- (N2);
        \draw (N2) -- (N3);
        \draw (N3) -- (N4);
        \draw (N4) -- (N1);
    \end{tikzpicture}
    \caption{}
\end{subfigure}
\caption{Three examples of degeneracy regions taken from Ref.~\onlinecite{penz2023geometry} in their respective tetrahedral and octahedral density domains, together with the corresponding lattice.
The lattice (a) is occupied by just one spinless particle, so the four vertices of the density domain are $(1,0,0,0)$ plus all permutations, while in (b) and (c) it is two spinless particles that lead to the six vertices $(1,1,0,0)$ plus all permutations. The shapes inside take up a large portion of the whole density domain and belong to degenerate ground states.}
\label{fig:intro}
\end{figure*}

And indeed, in the study of ground-state solutions on finite lattice systems without spin (at zero temperature), it was only recently discovered by the authors that already in very simple lattice systems counterexamples to the HK theorem exist~\cite{penz2021DFTgraphs}. 
For the spin-resolved version of density-functional theory (SDFT), on the other hand, such counterexamples have already been discussed before, both for the continuum~
\cite{Barth1972,capelle2001nonuniqueness,Eschrig2001} and on lattices~\cite{ullrich2005nonuniqueness-spin}.
So while there is representability of ground-state densities on the lattice by a corresponding choice of potentials, there is no \emph{unique} representability.
The phenomenon can be linked to the occurrence of a critical number of zero components in the wave-function expansion with respect to the lattice basis~\cite{penz2021DFTgraphs}. Interestingly, this again relates to degeneracy, since the potential can then be varied without affecting the ground state until a level crossing with an excited state is reached. This creates an intriguing geometrical picture that explains the HK counterexamples as the intersection points of degeneracy regions (themselves objects from algebraic geometry) in density space with each other or with the boundary of the density domain~\cite{penz2023geometry}. A few examples for degeneracy regions of simple lattice systems are given in Fig.~\ref{fig:intro}.
A further surprising discovery is the existence of $v$-representable densities even on the boundary of the density domain (i.e., with $\rho_i=0$ or $1$), exactly where a density region touches the boundary.
And even though degeneracy occurs only for very special potentials, either being highly symmetric or leading to an accidental degeneracy, the complex of all degeneracy regions can occupy a large part of the total density domain. As \citet{ullrich2002} expressed it in their pioneering work on topology in the potential and density spaces: ``Ground-state degeneracies in $v$ space are `rare' [while] degeneracy in $\rho$ space is \emph{not} `rare'.''
All this highlights the special relevance of degeneracy when discussing DFT, not only on lattices, and it offers exciting links to various fields of mathematics like topology, algebraic geometry, linear algebra, group theory, and graph theory.

In this perspective we aim at summarizing previous findings, especially those from Refs.~\onlinecite{penz2021DFTgraphs} and \onlinecite{penz2023geometry}, in a setting that explicitly includes the spin degree-of-freedom. We do this by first introducing the basic setting (Sec.~\ref{sec:setting}) and the density and potential spaces that form the basis of the further geometrical inquiries (Sec.~\ref{sec:spaces}). This sets the stage to explain the important notions of $v$-representability, non-uniqueness, and degeneracy regions and to show how they are all related (Sec.~\ref{sec:v-rep-deg-reg}). We are such able to give a purely geometrical formulation of the HK theorem. In order to assist the whole discussion, the different involved sets in density and potential space are introduced and explained separately (Sec.~\ref{sec:sets}). We continue with the very interesting example of an Anderson impurity model, where several of the before theoretically introduced phenomena can be found, and we describe their relation to the Kohn--Sham method (Sec.~\ref{sec:verdozzi-model}). 
Next to the ground-state case, we also discuss adiabatic variation of the potential and its connection to the geometric phase (Sec.~\ref{sec:adiabatic}), as well as the time-dependent setting (Sec.~\ref{sec:td}).
Finally, we summarize the remaining open questions and give an outlook towards possible future research directions (Sec.~\ref{sec:outlook}).

\section{Spin-lattice systems}
\label{sec:setting}

We consider putting $N$ spin-$\frac{1}{2}$ particles on $M>N/2$ sites. The corresponding one-particle Hilbert space is thus $\C^{2M}$ which gets promoted into the $N$-particle Hilbert space $\H = (\C^{2M})^{\wedge N}$ by an $N$-fold antisymmetric tensor product. Throughout this paper, we will remain in a \emph{finite-dimensional} Hilbert-space setting. The lattice indices will always be $i,j,k \in \{ 1,\ldots,M \}$ and the spin indices $\alpha,\beta \in \{ \up,\down \}$ with respect to $\hat\sigma_z$. We employ shorthand notation for index tuples, $i\alpha \equiv (i,\alpha)$.
We use the usual fermionic creation and annihilation operators $\aopdag_{i\alpha}, \aop_{i\alpha}$ of the spin-lattice basis and write $\rhoop_{i\alpha}=\aopdag_{i\alpha} \aop_{i\alpha}$ for the number (density) operator. If one index is left out, this implies summation over the other one, e.g., $\rhoop_{i} = \rhoop_{i\up} + \rhoop_{i\down}$ or $\rhoop_{\up} = \sum_i \rhoop_{i\up}$.
The Hamiltonian of the system will at first be allowed to be an entirely general self-adjoint operator on $\H$ (next to spin-lattice hopping it may include interactions between lattice sites, a Hubbard-$U$-type interaction between spin components, etc.), but we already write the action of an external potential and a magnetic field separately,
\begin{equation}\label{eq:ham-B}
    \hat H = \hat H_0 + \sum_i \left( v_i \rhoop_i + \BB_i\cdot\mop_i \right).
\end{equation}
Here, $\BB_i\in\R^3$ and the magnetization operator at $i$ is given by a 3-vector of operators
\begin{equation}
    \mop_i = (\aopdag_{i\up}, \aopdag_{i\down}) \cdot \ssigma \cdot \begin{pmatrix} \aop_{i\up} \\ \aop_{i\down} \end{pmatrix},
\end{equation}
with $\ssigma$ the 3-vector of Pauli matrices.
This can equivalently be rewritten with a complex 4-component potential at every lattice site,
\begin{equation}\label{eq:ham-4component}
    \hat H = \hat H_0 + \sum_{i \alpha\beta} v_{i \alpha\beta} \rhoop_{i \alpha\beta},
\end{equation}
where $v_{i \alpha\beta} = v_{i \beta\alpha}^*$ and we follow the standard definition for the density-matrix operator $\rhoop_{i \alpha\beta} = \aopdag_{i\beta}\aop_{i\alpha}$. Both conventions, scalar potential and magnetic field, or 4-component potential, are frequently employed. Yet, in the second formulation, the density-matrix operator $\rhoop_{i \alpha\beta}$ is not self-adjoint and yields complex expectation values, and also the potential $v_{i \alpha\beta}$ is complex, although their combined sum in Eq.~\eqref{eq:ham-4component} is again self-adjoint as it should.

In many cases one considers a \emph{collinear} setting, where the direction of the magnetic field is fixed at every lattice point and only the magnetization in direction of $\BB_i$ matters. Let the magnitude of the magnetic field be $B_i$ and the magnetization operator in the given direction $\hat m_i$, then the Hamiltonian of Eq.~\eqref{eq:ham-B} is reduced to
\begin{equation}\label{eq:ham-B-scalar}
    \hat H = \hat H_0 + \sum_i \left( v_i \rhoop_i + B_i\cdot\hat m_i \right).
\end{equation}
With $\rhoop_i = \rhoop_{i \up}+\rhoop_{i \down}$, $\hat m_i = \rhoop_{i \up}-\rhoop_{i \down}$ and $v_{i\up} = v_i+B_i$, $v_{i\down} = v_i-B_i$ we arrive at the more usually applied formulation for the collinear setting, where two potentials act individually on the spin components,
\begin{equation}\label{eq:ham-collinear}
    \hat H = \hat H_0 + \sum_{i \alpha} v_{i \alpha} \rhoop_{i \alpha}.
\end{equation}
Finally, if the magnetic field is entirely removed or taken as fixed and integrated into $\hat H_0$, one arrives at the usual DFT setting, where the spin components cannot be addressed individually by the external potential,
\begin{equation}\label{eq:ham-usual-DFT}
    \hat H = \hat H_0 + \sum_{i} v_{i} \rhoop_{i}.
\end{equation}

In all those settings we can speak of a \emph{graph} instead of a lattice, since we are concerned with how spin-lattice points are connected by $\hat H$ and since this can happen in an irregular way. For $i\alpha \neq j\beta$, we say that $i\alpha$ and $j\beta$ are connected (or \emph{adjacent}), written $i\alpha \sim j\beta$, if $\hat H$ includes any terms with $\aopdag_{i\alpha}\aop_{j\beta}$ (self-adjointness dictates that it then also needs to include the adjoint terms).
If the term is $\aopdag_{i\alpha}\aop_{j\alpha}$, this is called `hopping', while a Hubbard-$U$ interaction term connects different spin values, 
\begin{equation}\begin{aligned}
    \rhoop_{i\uparrow}\rhoop_{i\downarrow} &= \aopdag_{i\uparrow} \aop_{i\uparrow}\aopdag_{i\downarrow} \aop_{i\downarrow} = (1-\aop_{i\uparrow}\aopdag_{i\uparrow}) \aopdag_{i\downarrow} \aop_{i\downarrow} \\
    &= \rhoop_{i\downarrow} - (\aopdag_{i\downarrow}\aop_{i\uparrow})(\aopdag_{i\uparrow}\aop_{i\downarrow}).
\end{aligned}\end{equation}
If this graph of spin-lattice points decomposes into several connected components then the Hamiltonian has a block-diagonal form and we actually speak about totally separated systems that can be dealt with individually.
Such a situation automatically leads to degeneracy and the ensuing non-uniqueness that will be discussed in Sec.~\ref{sec:v-rep-deg-reg}.
Remember that in some cases a block-diagonal form can be achieved by unitary transformation of the Hamiltonian, like in the example discussed in Sec.~\ref{sec:verdozzi-model}. This means that by symmetry one is able to pass to a new basis and find new creation and annihilation operators (like in a Bogoliubov transformation leading to a \emph{quasiparticle} interpretation), such that the system decouples into independent subsystems which will decrease computational complexity. Systematically, this is performed by finding the \emph{irreducible representations} of the symmetry group. Yet, note that by allowing an independent potential at every spin-lattice point, any form of symmetry will be broken.

As a trivial example of disconnected graphs, imagine that $\hat H_0$ in Eqs.~\eqref{eq:ham-collinear} or \eqref{eq:ham-usual-DFT} does not connect the up/down spin components. Since the $v_{i\alpha}\hat\rho_{i\alpha}$ cannot provide such a connection either, the total Hamiltonian already decomposes into separate up- and down-spin systems. If this situation is still considered as a single system then it directly leads to (trivial) HK counterexamples~\cite{capelle2001nonuniqueness,ullrich2005nonuniqueness-spin}.

\section{Density and potential spaces}
\label{sec:spaces}

As soon as we move our attention to DFT, we need to choose certain observables that serve as density variables and that couple linearly to the external potentials. According to Sec.~\ref{sec:setting}, three choices offer themselves.
\begin{center}
\begin{tabular}{l|l|l}
  & density & potential \\ \hline
 standard DFT & $\rho_i$ & $v_i$ \\ \hline
 collinear SDFT & $\rho_{i\alpha}$ & $v_{i\alpha}$ \\  
  & $\rho_i,m_i$ & $v_i,B_i$ \\ \hline
 non-collinear SDFT & $\rho_{i\alpha\beta}$ & $v_{i\alpha\beta}$ \\
 & $\rho_i,\mm_i$ & $v_i,\BB_i$
\end{tabular}
\end{center}
Here, the density values themselves are always considered the expectation values of the corresponding operators. This can either be with respect to a pure state, $\rho_i = \langle \rhoop_i \rangle_\Psi = \langle \Psi,\rhoop_i\Psi \rangle$, or a mixed (ensemble) state, $\rho_i = \langle \rhoop_i \rangle_\Gamma = \Tr ( \Gamma\rhoop_i )$, and accordingly for the other density variables. Note that a totally general and more abstract formulation employing a set of $M$ self-adjoint `density' operators $\hat o_i$, $i\in \{1,\ldots,M\}$, that are linearly coupled to `potentials' $v_i\in\R$,
\begin{equation}
    \hat H = \hat H_0 + \sum_i v_i \hat o_i,
\end{equation}
is possible as well~\cite{schonhammer1995density,Song2023}.

The usual density as site occupancy naturally allows $\rho_i\in[0,2]$ due to the antisymmetry of the wave function and the possibility of filling two different spin channels. In the spin-resolved formulation this changes to $\rho_{i\alpha}\in[0,1]$. Additionally, we have the normalization to the number of particles $N$,
\begin{equation}
    \sum_i\rho_i=\sum_{i\alpha}\rho_{i\alpha}=N.
\end{equation}
We thus define the density set
\begin{equation}
    \Dens^{\mathsf{DFT}} = \left\{ \rho \in \R^{M} \mid 0\leq \rho_{i} \leq 2, \sum\nolimits_{i} \rho_{i}=N \right\}
\end{equation}
for standard lattice DFT with spin-$\frac{1}{2}$ particles. This results in a doubly scaled $(M,N/2)$-hypersimplex for $\rho_{i}$ if $N$ even. 
If $N$ odd the density space for $\rho_{i}$ is slightly more complex, but it is always described by a $(M-1)$-dimensional convex polytope.
If the particles are assumed spinless instead, the restricting inequality is $0\leq \rho_{i} \leq 1$ and the resulting shape is a $(M,N)$-hypersimplex. \cite{penz2021DFTgraphs,Rispoli2008-hypersimplex}
In the case of collinear SDFT we get
\begin{equation}
    \Dens^{\mathsf{cSDFT}} = \left\{ \rho \in \R^{2M} \mid 0\leq \rho_{i\alpha} \leq 1, \sum\nolimits_{i\alpha} \rho_{i\alpha}=N \right\}.
\end{equation}
This results in a $(2M,N)$-hypersimplex for $\rho_{i\alpha}$ if $N$ even and a more general $(2M-1)$-dimensional convex polytope if $N$ odd.

To determine the space of possible magnetizations $\mm_i = \langle \mop_i \rangle$ for non-collinear SDFT, let us first consider just a single particle on a single vertex and the expectation value with respect to a pure state $\Psi\in\C^2$. We have
\begin{equation}
    |\mm|^2 = |\langle \hat\ssigma \rangle_\Psi|^2 = \langle \hat\sigma_x \rangle_\Psi^2+\langle \hat\sigma_y \rangle_\Psi^2+\langle \hat\sigma_z \rangle_\Psi^2 = 1,
\end{equation}
which means the magnetization vector $\mm$ is always on a sphere. For mixed states convex combinations of the magnetization vectors become possible, so the value can also be inside the sphere and $|\mm|\leq 1$. We recognize exactly the Bloch sphere construction, where pure states are located on the surface of the sphere and mixed states are always in the interior. The space for magnetization is thus the unit ball, $\mm\in\B$.
Adding another particle on the same site will force them to align antiparallel with a total magnetization of zero, so this does not change the space.
For multiple vertices, the value of $\mm_i = \langle \mop_i \rangle_\Psi$ depends on the partial-trace density matrix over all vertices $j\neq i$, so one already considers a mixed state and thus, in general, has $|\mm_i|\leq 1$. The total space for magnetization is thus $\B^M$. In the collinear setting this accordingly changes to $[-1,1]^M$.
Whatever setting we use, the complete density space for $(\rho_i,\mm_i)$, $(\rho_i,m_i)$, $\rho_{i\alpha}$, or just $\rho_i$ will always be denoted by $\Dens$.
It was previously noted in the context of standard lattice DFT that for any $\rho\in\Dens$ there is indeed a $\Psi\in\H$ that gives this density~\cite[Prop.~5]{penz2021DFTgraphs}. This result can easily be extended to both variants of SDFT.

For the following presentation we will mostly remain in the setting of \emph{collinear SDFT}, while statements for the other DFT variants follow in a similar manner. We write $\hat H_v$ for the Hamiltonian from Eq.~\eqref{eq:ham-collinear} that includes a potential $v_{i\alpha}$.
This potential, according to Eq.~\eqref{eq:ham-collinear}, acts in the form of an inner product on $\R^{2M}$ when the expectation value is formed and we write
\begin{equation}
    \langle \hat H_v \rangle_\Psi = \langle \hat H_0 \rangle_\Psi + \sum_{i \alpha} v_{i \alpha} \rho_{i \alpha} = \langle \hat H_0 \rangle_\Psi + \langle v,\rho \rangle.\end{equation}
In this sense it is natural to see the potentials as dual to densities. But $\Dens$ lies within an affine space of codimension 1, defined by the normalization condition $\sum_{i\alpha} \rho_{i\alpha}=N$. Now, if two potentials differ by just a constant, $v-v'=c$, then their difference acts as $\langle v-v',\rho\rangle = cN$ on any $\rho\in\Dens$. This means that $v,v'$ cannot be distinguished as elements of the dual space $\Pot = \Dens^*$. In other words, the elements $[v]\in\Pot$ are equivalence classes where for all $v,v'\in [v]$ it always holds that $v-v'$ is constant. In the following, we will denote potentials further simply as $v\in\Pot$ and understand them as modulo a constant. This means that the usual statement in DFT that potentials are defined ``up to a constant'', since adding a constant does not change the ground-state properties (it only shifts the ground-state energy), can actually be suppressed. For each $v\in\Pot$ we now define the ground-state energy functional
\begin{equation}
    E(v) = \inf_{\Psi} \langle \hat H_v \rangle_\Psi,
\end{equation}
where the variation extends over all $\Psi\in\H$ normalized to 1. This functional is concave in $v$ because of linearity in $v$ and the properties of the infimum. Another immediate result is that this infimum is realized by some (or rather many) $\Psi\in\H$ since the variation domain is compact in the case of a finite lattice. One could thus equally write `min' instead of `inf'. The respective optimizer is then a ground-state wave function for $\hat H_v$.

If for a given $\rho\in\Dens$ there is further a potential $v\in\Pot$ such that the density is achieved by the corresponding (possibly ensemble) ground state of $\hat H_v$ then one calls this density \emph{$v$-representable}. This is a subtle notion, but it can be proven that all densities from the interior of $\Dens$ are $v$-representable, while only a certain few densities on the boundary of $\Dens$ have this property~\cite{CCR1985,penz2021DFTgraphs,penz2023geometry}. Importantly, this notion depends on the choice of $\hat H_0$ that then also determines which boundary densities can be reached. We will give a full characterization of $v$-representable densities in the spin-lattice setting in Sec.~\ref{sec:v-rep-deg-reg} below. One has to keep in mind that the situation is rather different for the infinite-dimensional continuum setting of standard DFT, where $v$-representability depends on the topology of the density space and only limited results are available by now \cite{Lammert2007,lammert2006coarse,lammert2010well,sutter2023solution}.

\section{\texorpdfstring{$v$}{v}-representability, non-uniqueness, and degeneracy regions}
\label{sec:v-rep-deg-reg}

One of the cornerstones in the presentation of DFT is usually the HK theorem that establishes a mapping from $v$-representable densities to potentials. In its usual formulation it states that for a fixed $\hat H_0$ and every $\rho\in\Dens$ that is $v$-representable, a \emph{unique} $v\in\Pot$ can be found that retrieves this density from a ground state. This can be an ensemble state (represented by a density matrix) in the case of ground-state degeneracy. In Sec.~\ref{sec:verdozzi-model} we will discuss an example where the occurrence of degeneracy was mistakenly taken as a failure of $v$-representability.
But there are counterexamples (discussed at the end of this section) that rule out the full validity of this statement and show the \emph{non-uniqueness} of the representing potential. Nevertheless, a characterization similar to that of $v$-representability can also be found when such counterexamples arise and this will lead us to an alternative and entirely geometrical formulation for the HK theorem.

The theoretical basis for DFT is the convex \emph{universal density functional} that can be defined on $\Dens$ as the density-matrix constrained-search functional~\cite{Lieb1983}
\begin{equation}\label{eq:F-def}
    F(\rho) = \inf_{\Gamma\mapsto\rho} \Tr \hat H_0 \Gamma.
\end{equation}
Here, variation is over all density matrices that yield the given density $\rho$. The ground-state energy for a given $v\in\Pot$ is then
\begin{equation}\label{eq:E-from-F}
    E(v) = \inf_{\Gamma} \Tr \hat H_v \Gamma = \inf_{\rho\in\Dens}\left\{ F(\rho) + \langle v,\rho \rangle \right\}.
\end{equation}
In other words, $E(v)$ is the Legendre--Fenchel transform (with non-standard sign convention) of $F(\rho)$ and concave as such. The back-transformation then gives $F(\rho)$ again,
\begin{equation}\label{eq:F-LF}
    F(\rho) = \sup_{v\in\Pot}\left\{ E(v) - \langle v,\rho \rangle \right\}.
\end{equation}
More on these functionals can be found in any mathematically oriented introduction to DFT~\cite{engel-dreizler-book,Penz2023-DFT-Review-Part-1}.
The optimum in Eqs.~\eqref{eq:E-from-F} and \eqref{eq:F-LF} is attained where the (concave or convex) functionals allow a zero tangent functional. Here, a tangent functional to a convex functional $f$ at $\rho\in\Dens$ is any $v\in\Pot$ such that $f(\rho')\geq f(\rho) + \langle v,\rho'-\rho\rangle$ for all $\rho'\in\Dens$. For concave functionals the inequality is reversed. Further note that if $f$ is non-differentiable at $\rho$, i.e., it has a kink, the tangent functional is non-unique. The set of all tangent functionals to a convex functional $f$ is called the subdifferential, written $\underline\partial f(\rho)$. With $f(\rho) = F(\rho) + \langle v,\rho \rangle$ we have
\begin{equation}\label{eq:F-subdiff}
    0 \in \underline\partial f(\rho) = \underline\partial F(\rho) + v \quad\Leftrightarrow\quad -v \in \underline\partial F(\rho)
\end{equation}
as a condition for $v\in\Pot$ being a maximizer in Eq.~\eqref{eq:F-LF} and thus fulfilling $F(\rho)=E(v)-\langle v,\rho \rangle$. But this conversely implies $E(v)=F(\rho)+\langle v,\rho \rangle$ and so this $\rho\in\Dens$ is the minimizer in Eq.~\eqref{eq:E-from-F} and equivalently an element in the superdifferential of the concave $E(v)$,
\begin{equation}
    \rho \in \overline\partial E(v).
\end{equation}
So the above relation tells us that $\rho$ is a ground-state density for $v$. The ground-state density can be non-unique if degeneracy occurs for the chosen $v$ and this phenomenon will lead us to the next, important definition.

We define a \emph{degeneracy region} $D(v) \subseteq \Dens$ as the set of densities coming from all (ensemble) ground states of the Hamiltonian $\hat H_v$ with a $v\in\Pot$ that facilitates ground-state degeneracy.
By what has been said above, it is equal to the superdifferential of the concave ground-state energy functional $E(v)$,
\begin{equation}\begin{aligned}
    D(v) &=\{\rho\in\Dens\mid \Gamma\mapsto\rho, \Tr \hat H_v \Gamma=E(v) \} \\&=\overline{\partial}E(v) = \{\rho\in\Dens\mid \forall v'\in\Pot:\\&\qquad\qquad\qquad E(v')\leq E(v)+\langle v'-v,\rho\rangle\}.
\end{aligned}\end{equation}
A density region consists of $v$-representable densities (by definition) and it is always convex and closed.
The maximal dimension of a degeneracy region when the degeneracy is $g$-fold is $g^2-1$, reduced to $g(g+1)/2-1$ in the case of a real $\hat H_0$. When the density variable does not contain any spin information, degeneracy can be such that it only affects the internal spin degree-of-freedom and is thus not expressed in the density alone. Then $D(v)$ remains a single point and we call the degeneracy `internal'. This is resolved in SDFT with the inclusion of the magnetization as an additional density variable. Then, full spin degeneracy means that $D(v)$ extends over all of $\B^M$.

\begin{figure}[ht]
  \begin{subfigure}{.45\columnwidth}
  \centering
    \begin{tikzpicture}[scale=0.45]
        \draw[-,opacity=0] (4.5, 4.5) -- (4.5, 8.5); 
        \fill[red,opacity=0.15] (1.5, 6.5) circle (1.5);
        \fill[red,opacity=0.15] (4.5, 6.5) circle (1.5);
        \draw[red] (1.5, 6.5) circle (1.5); 
        \draw[red] (4.5, 6.5) circle (1.5);
        \fill[MidnightBlue] (3, 6.5) circle (4pt);
    \end{tikzpicture}
    \caption{}
  \end{subfigure}
  \hfill 
  \begin{subfigure}{.45\columnwidth}
    \centering
    \begin{tikzpicture}[scale=0.45]
        \pattern[pattern=north east lines] (4.5, 4.5)--(4.8, 4.5)--(4.8, 8.5)--(4.5, 8.5)--cycle;
        \draw[-, thick] (4.5, 4.5) -- (4.5, 8.5);
        \fill[red,opacity=0.15] (3, 6.5) circle (1.5);
        \draw[red] (3, 6.5) circle (1.5); 
        \fill[MidnightBlue] (4.5, 6.5) circle (4pt);
    \end{tikzpicture}
    \caption{}
  \end{subfigure}
  \caption{Degeneracy regions that intersect or touch the boundary of the density domain.}
  \label{fig:deg-touch}
\end{figure}
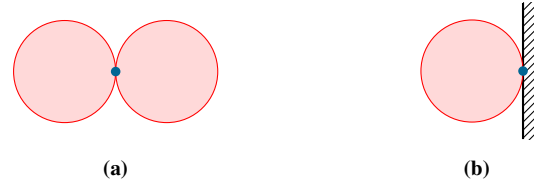

\begin{figure}[ht]
  \begin{subfigure}{.45\columnwidth}
  \centering
    \begin{tikzpicture}[scale=0.45]
        \draw[->] (0, 0) -- (6.5, 0) node[below] {$\lambda$};
        \draw[->] (0, 0) -- (0, 4) node[left] {$E$};
        \draw[domain=0:1.6, smooth, very thick, variable=\x, MidnightBlue] plot ({\x}, {-1/3*(\x-3)*(\x-3)+3.25});
        \draw[domain=1.6:4.4, smooth, variable=\x, MidnightBlue] plot ({\x}, {-1/3*(\x-3)*(\x-3)+3.25});
        \draw[domain=4.4:6, smooth, very thick, variable=\x, MidnightBlue] plot ({\x}, {-1/3*(\x-3)*(\x-3)+3.25});
        \draw[-, MidnightBlue] (0, 2.5) -- (1.5, 2.5);
        \draw[-, MidnightBlue, very thick] (1.5, 2.5) -- (4.5, 2.5);
        \draw[-, MidnightBlue] (4.5, 2.5) -- (6, 2.5);
        \node[below] at (1.5,0) {$\lambda_1$};
        \node[below] at (4.5,0) {$\lambda_2$};
        \draw[-,dashed] (1.5, 0) -- (1.5, 5);
        \draw[-,dashed] (4.5, 0) -- (4.5, 5);
        \fill[red] (1.5, 2.5) circle (4pt);
        \fill[red] (4.5, 2.5) circle (4pt);
        \fill[red,opacity=0.15] (1.5, 6.5) circle (1.5);
        \fill[red,opacity=0.15] (4.5, 6.5) circle (1.5);
        \draw[red] (1.5, 6.5) circle (1.5); 
        \draw[red] (4.5, 6.5) circle (1.5);
        \fill[MidnightBlue] (3, 6.5) circle (4pt);
        \draw[domain=0:1, smooth, thick, variable=\x, MidnightBlue] plot ({-\x}, {0.2*\x*\x+6.5});
        \draw[domain=0:1, smooth, thick, variable=\x, MidnightBlue] plot ({\x+6}, {-0.2*\x*\x+6.5});
    \end{tikzpicture}
    \caption{}
  \end{subfigure}
  \hfill 
  \begin{subfigure}{.45\columnwidth}
    \centering
    \begin{tikzpicture}[scale=0.45]
        \draw[->] (0, 0) -- (6.5, 0) node[below] {$\lambda$};
        \draw[->] (0, 0) -- (0, 4) node[left] {$E$};
        \draw[-, MidnightBlue] (0, 2) -- (3, 2);
        \draw[->, MidnightBlue, very thick] (3, 2) -- (6, 2) node[right] {$\infty$};
        \draw[-, MidnightBlue, very thick] (0, 0.5) -- (3, 2);
        \draw[-, MidnightBlue] (3, 2) -- (6, 3.5);
        \draw[-,dashed] (3, 0) -- (3, 5);
        \fill[red] (3, 2) circle (4pt);
        \draw[->, thick] (2.2, -.5) to (2.9, -.1);
        \node[below,left] at (2.3, -.6) {$v_{\rm A}$};
        \pattern[pattern=north east lines] (4.5, 4.5)--(4.8, 4.5)--(4.8, 8.5)--(4.5, 8.5)--cycle;
        \draw[-, thick] (4.5, 4.5) -- (4.5, 8.5);
        \fill[red,opacity=0.15] (3, 6.5) circle (1.5);
        \draw[red] (3, 6.5) circle (1.5); 
        \fill[MidnightBlue] (4.5, 6.5) circle (4pt);
        \draw[domain=0:1, smooth, thick, variable=\x, MidnightBlue] plot ({-\x+1.5}, {-0.2*\x*\x+6.5});
    \end{tikzpicture}
    \caption{}
  \end{subfigure}
  \caption{The two situations of degeneracy regions intersecting and touching the boundary related to energy level crossings.}
  \label{fig:crossing}
\end{figure}
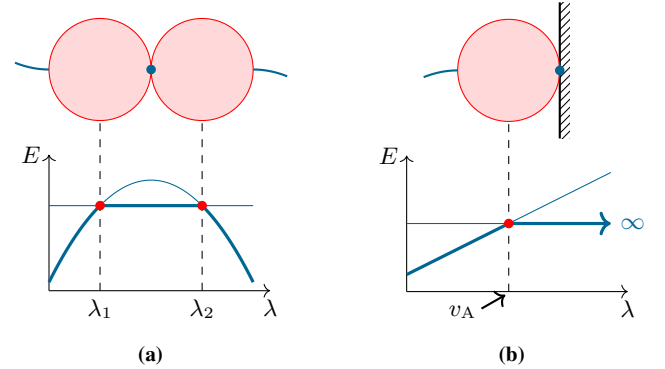

When degeneracy is due to symmetry and a variation of $v$ preserves this symmetry, then this does not change the degeneracy of the ground state, although the associated density region $D(v)$ of course changes. This way, a family of potentials $v(\lambda)$, continuously parametrized by a vector $\lambda\in\R^\ell$, can lead to a family of degeneracy regions $D(v(\lambda))$ that we call a \emph{degeneracy bundle}. It consists of degeneracy regions that lie arbitrarily close together but never touch. Since the space of densities has dimension $2M-1$ (in the collinear SDFT setting), we get a maximum of $\ell\leq 2M-d-1$ parameters for a family of degeneracy regions with dimension $d$ each. This reproduces an earlier result by \citet{ullrich2002} that was later adapted to degeneracy regions~\cite{penz2023geometry}. But degeneracy regions, if they are not part of the same bundle, can intersect and it is also possible that they touch the boundary of the density domain itself. These two situations are sketched in Fig.~\ref{fig:deg-touch} and they have a special relevance in the current discussion, since there are always infinitely many different potentials that yield those density points as ground-state densities.
They are thus counterexamples to the HK theorem and must be omitted in its formulation (see below). How this situation arises can indeed be easily recognized. If a density is produced by two different potentials $v$ and $v'$ then the linearity of the Hamiltonian means that any linear combination of those potentials leads to the same density, \emph{unless} an energy-level crossing is reached in either direction. This then gives rise to the two possible situations, pictorially related to energy-level crossings in Fig.~\ref{fig:crossing}.

Finally, we need to highlight that even though we presented these notions within a specific spin-lattice DFT setting, they hold quite generally, even for the infinite-dimensional continuum setting. Details about the proof and the intricate shape of density regions can be found in our study on the geometry of degeneracy~\cite{penz2023geometry}.
We here summarize the results:

\begin{enumerate}[(a)]
    \item\label{HK-item-intersect} If two density regions $D(v)$ and $D(v')$ intersect then $D(v)\cap D(v')$ is either a single ground-state density point or a density region itself. In both cases it results from all potentials that are a convex combination of $v$ and $v'$.

    \item\label{HK-item-touching-boundary} If a density region $D(v)$ touches the boundary of $\Dens$ then this density point results from all potentials that lie on a ray that extends from $v$ to infinity.

    \item\label{HK-item-vrep} All densities that are not on the boundary of $\Dens$ are $v$-representable. Densities on the boundary of $\Dens$ need to be part of a degeneracy region in order to be $v$-representable. All $v$-representable densities that are not described by \ref{HK-item-intersect} or \ref{HK-item-touching-boundary} are even \emph{uniquely} $v$-representable.
    
\end{enumerate}

We are thus in the position to formulate a purely geometrical HK theorem: \emph{All ground-state densities that are not on the boundary of the density domain and that are not at the intersection of degeneracy regions are uniquely given by a potential.}\\

Examples for all mentioned situations \ref{HK-item-intersect}-\ref{HK-item-vrep} in the pure-lattice case can be found in our previous works on the topic~\cite{penz2021DFTgraphs,penz2023geometry} and seemed to have been overlooked previously as counterexamples to the HK theorem.

We want to conclude this section with a brief discussion of the mentioned HK counterexamples that specifically connect to spin in the collinear setting~\cite{capelle2001nonuniqueness,ullrich2005nonuniqueness-spin}.
In those examples the up- and down-components of the spin lattice remain unconnected by the Hamiltonian (in the sense given in Sec.~\ref{sec:setting}) and consequently changing the potential $v_{i\alpha}$ by adding a constant for all $i$ on only one component $\alpha\in\{\up,\down\}$ does not influence the ground state, \emph{unless} this change leads to a level crossing. But the occurrence of a level crossing means that the density is part of a degeneracy region and a change of the potential in the other direction either leads to another level crossing (\ref{HK-item-intersect}, intersecting degeneracy regions) or it continues to infinity without a level crossing (\ref{HK-item-touching-boundary}, touching the boundary of the density domain at full magnetization $m_i \in \{\pm 1\}$). The non-collinear setting for a single lattice vertex is instructive too. Here, at $\BB=0$, the whole Bloch sphere is a degeneracy region since no spin-direction is preferred. Since the density domain is just the same Bloch sphere, the degeneracy region touches the boundary at every $\mm$ with $|\mm|=1$ (pure states) and indeed all fields $\BB\neq 0$ lead to non-uniqueness in the sense of \ref{HK-item-touching-boundary}.

\section{Sets of the potential-density mapping}
\label{sec:sets}

We start by defining the set $\PotDeg$ of all potentials that lead to degeneracy in the ground state and the complementary set $\PotNonDeg$ where no degeneracy occurs. Obviously, the complete space of potentials is the disjoint union of those sets, $\Pot = \PotNonDeg \;\dot{\cup}\; \PotDeg$. The set $\PotDeg$ is closed since degeneracy occurs only when (continuous) energy bands intersect, which makes $\PotNonDeg$ open. Now define the mapping from potentials to ground-state densities that we just call $\rho$ for convenience.
\begin{equation}\label{eq:rho-map}
    \rho : \PotNonDeg \longrightarrow \Dens
\end{equation}
It is well-defined on $\PotNonDeg$ since each such potential leads to a \emph{unique} ground-state density due to the lack of degeneracy. A beautiful result of \citet{rellich1937,rellich-book} then tells us that this mapping is analytic, so in particular it is continuous. We also define the image of this map, $\DensNonDeg = \rho(\PotNonDeg)$, which is found to be open~\cite{KohnPRL,penz2021DFTgraphs}, and further $\DensDeg$ as the union of all degeneracy regions. In many examples the occurrence of degeneracy in potential space is rare ($\PotDeg$ has measure zero), while the corresponding densities $\DensDeg$ can fill a larger volume than their complement $\Dens\setminus\DensDeg$~\cite{ullrich2002,penz2023geometry}.

It is now important to note that the sets $\DensDeg$ and $\DensNonDeg$ are \emph{not} necessarily disjoint. Take the set $\DensNonDeg\cap\DensDeg$, then this includes densities that come from different potentials, exactly counterexamples to the HK from before.
If we call $\DensNonUv$ the set of densities that arise from multiple different potentials (also called `non-uv densities' after 'non-uniquely $v$-representable' in Ref.~\onlinecite{penz2021DFTgraphs}), then clearly $\DensNonDeg\cap\DensDeg\subseteq\DensNonUv$. Sec.~\ref{sec:v-rep-deg-reg}~\ref{HK-item-intersect}-\ref{HK-item-vrep} gives a complete classification of $\DensNonUv$ and also has $\DensNonUv\subseteq\DensDeg$, since all those counterexamples are connected to degeneracy.
It is known that $\DensNonUv$ is closed and has measure zero in $\Dens$~\cite[Cor.~10]{penz2023geometry} and that densities in $\DensNonUv$ can only arise from wave functions with a certain minimal number of zero coefficients in the lattice basis~\cite[Sec.~III.B]{penz2021DFTgraphs}.
Then there is the set $\DensNonDeg\cup\DensDeg$ that collects all $v$-representable densities.
In general it holds $\DensNonDeg\cup\DensDeg \subsetneqq \Dens$, since there can always be densities on the boundary of $\Dens$ that are not $v$-representable.

All densities that come from multiple potentials are collected in $\DensNonUv$, so we can invert the mapping Eq.~\eqref{eq:rho-map} on $\DensNonDeg\setminus\DensNonUv = \DensNonDeg\setminus\DensDeg$. But what about continuity of the inverse map
$\rho^{-1}: \DensNonDeg\setminus\DensNonUv \to \Pot$.
This is always secured only if we already limit ourselves to a \emph{compact} domain in Eq.~\eqref{eq:rho-map}, which $\PotNonDeg$ is not since it is open.
But using tools from convex analysis we can still show that the inverse map is continuous almost everywhere too and even extend it. First, remember from Eq.~\eqref{eq:F-subdiff} that if $\rho$ is a ground-state density for potential $v$ then $-v$ is an element of the subdifferential of the universal density functional $F$ at $\rho$, $-v\in\underline\partial F(\rho)$. Since for $\rho\in (\DensNonDeg\cup\DensDeg)\setminus\DensNonUv$ this subdifferential contains only a single element, $F$ is differentiable at $\rho$ and we write $v=-\nabla F(\rho)$ \cite[Th.~25.1]{rockafellar-book}. 
Differentiability in this context means Fréchet differentiability, which in turn means that the inverse map 
\begin{equation}\begin{aligned}
    \rho^{-1}: (\DensNonDeg\cup\DensDeg)\setminus\DensNonUv &\to \Pot
    \\
    \rho &\mapsto -\nabla F(\rho),
\end{aligned}\end{equation}
defined on the \emph{uniquely $v$-representable} densities, is continuous \cite[Th.~25.5]{rockafellar-book}.
Actually this is more than an inverse to Eq.~\eqref{eq:rho-map} because it now includes degeneracy regions but it still does not include any boundary densities.
The domain of the inverse mapping above is exactly the one from the geometrical formulation of the HK theorem from Sec.~\ref{sec:v-rep-deg-reg}, ``all ground-state densities that are not on the boundary of the density domain and that are not at the intersection of degeneracy regions'', because those densities arise from multiple potentials.
Yet we are not allowed to continue the inverse mapping to all $v$-representable densities $\DensNonDeg\cup\DensDeg$ (this would require uniform continuity).
This means that a perfectly smooth density path can start to diverge if mapped back to potential space as one gets close to $\DensNonUv$.
Note that this is different to the discontinuity of $\rho^{-1}$ in the usual continuum setting~\cite{garrigue2021properties-pot-to-gs}, which depends on much more complicated topologies and where due to the HK theorem no non-uniquely $v$-representable densities are expected to exist.

Densities with zero components are only found on the boundary of $\Dens$, but the only $v$-representable densities on the boundary are in $\DensNonUv$ as stated in Sec.~\ref{sec:v-rep-deg-reg}. Conversely, all ground-state densities in $(\DensNonDeg\cup\DensDeg)\setminus\DensNonUv$, which removes the boundary densities, are strictly positive.

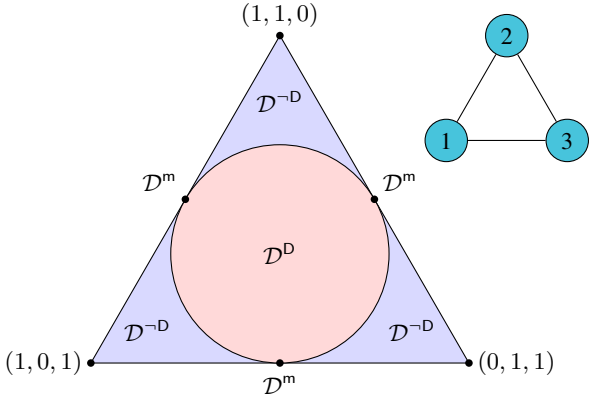
\begin{figure}[ht]
	\centering
	\begin{tikzpicture}
	\def\ORX{0}; 
	\def\ORY{-1};
	\def\L{5}; 
    \coordinate (R1) at (\ORX,{\ORY+\L*sqrt(3)/2-\L/2/sqrt(3)}); 
    \coordinate (R2) at ({\ORX-\L/2},{\ORY-\L/2/sqrt(3)}); 
    \coordinate (R3) at ({\ORX+\L/2},{\ORY-\L/2/sqrt(3)}); 
	\coordinate (RI) at ({\ORX-\L/4},{\ORY+\L*sqrt(3)/2/2-\L/2/sqrt(3)});
	\coordinate (RII) at ({\ORX+\L/4},{\ORY+\L*sqrt(3)/2/2-\L/2/sqrt(3)});
	\coordinate (RIII) at (\ORX,{\ORY-\L/2/sqrt(3)});
	\coordinate (RB) at ({\ORX-\L/4},{\ORY-\L/4/sqrt(3)});
    \path[fill=blue!15] (R1) -- (R2) -- (R3) -- (R1);
    \draw (R1) -- (R2) -- (R3) -- (R1);
    \node at (R1) [circle,fill,inner sep=1.0pt] {};
    \node at (R2) [circle,fill,inner sep=1.0pt] {};
    \node at (R3) [circle,fill,inner sep=1.0pt] {};
    \node[above] at (R1) {$(1,1,0)$};
    \node[left] at (R2) {$(1,0,1)$};
    \node[right] at (R3) {$(0,1,1)$};
	\draw [fill=red!15] (\ORX,\ORY) circle ({\L/2/sqrt(3)}); 
	\node [above left] at (RI) {$\DensNonUv$};
	\node at (RI) [circle,fill,inner sep=1.0pt] {};
	\node [above right] at (RII) {$\DensNonUv$};
	\node at (RII) [circle,fill,inner sep=1.0pt] {};
	\node [below=0.2em] at (RIII) {$\DensNonUv$};
	\node at (RIII) [circle,fill,inner sep=1.0pt] {};
	\node at (\ORX,\ORY) {$\DensDeg$};
	\node at ({(\ORX-\L/2)*0.7+\ORX*0.3},{(\ORY-\L/2/sqrt(3))*0.7+\ORY*0.3}) {$\DensNonDeg$};
	\node at ({(\ORX+\L/2)*0.7+\ORX*0.3},{(\ORY-\L/2/sqrt(3))*0.7+\ORY*0.3}) {$\DensNonDeg$};
	\node at ({\ORX*0.7+\ORX*0.3},{(\ORY+\L*sqrt(3)/2-\L/2/sqrt(3))*0.7+\ORY*0.3}) {$\DensNonDeg$};
    \def\CX{3}; 
    \def\CY{0.5};
    \def\L{0.8}; 
    \coordinate (C1) at (-1*\L+\CX,0+\CY);
    \coordinate (C2) at (0+\CX,{sqrt(3)*\L+\CY});
    \coordinate (C3) at (1*\L+\CX,0+\CY);
    \foreach \m in {1,...,3}
    	\node[graphnode] (N\m) at (C\m) {\m};
    \draw (N1) -- (N2) -- (N3) -- (N1);
	\end{tikzpicture}
	\caption{Different density sets for two spinless particles on the triangle graph. The three vertices are the extremal fillings $(1,1,0)$ plus all permutations. The area $\DensDeg$, a closed circular set, are all densities from degenerate ground states that occur when the potential is constant. The remaining three disconnected sets that form $\DensNonDeg$ are all open and belong to non-degenerate ground states from other potentials. Three non-unique $v$-representable points belong to $\DensNonUv$ and lie exactly at the intersections of the incircle with the triangle.}
	\label{fig:triangle-example}
\end{figure}

We exemplify these sets with the triangle-graph example ($M=3,N=2$) from Ref.~\onlinecite[Sec.~VI.C]{penz2021DFTgraphs}, depicted in Fig.~\ref{fig:triangle-example}.

\section{Anderson impurity
model: mistaken non-\texorpdfstring{$v$}{v}-representability and real convergence issues}
\label{sec:verdozzi-model}

We learnt in Sec.~\ref{sec:v-rep-deg-reg} that every density (that is not on the boundary) in a spin-lattice model is $v$-representable; still \citet{RoesslerVerdozzi2018} reported that they found a counterexample in exactly such a setting. Their claim rests on a numerical procedure that aims at reaching the density of an interacting system by adjusting the potential in a non-interacting system and that failed. The statement of non-$v$-representability is \emph{wrong}, but analyzing the situation that leads to this failure gives interesting insight into the relevance of degeneracy, especially with respect to Kohn--Sham DFT.

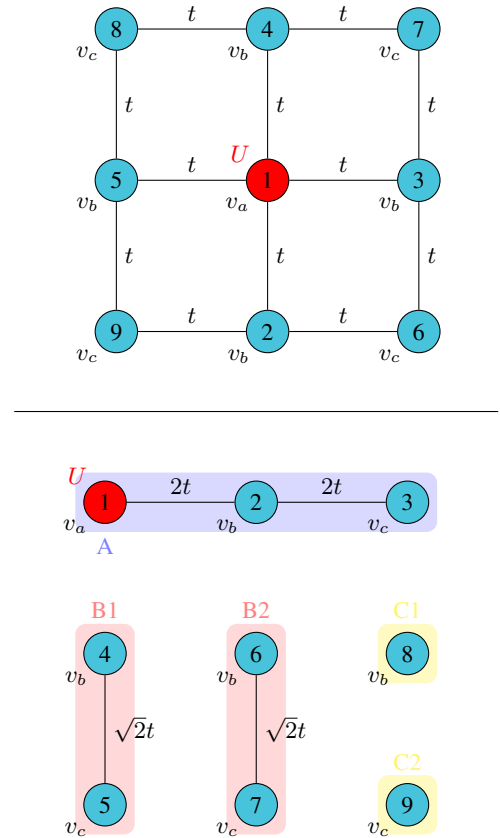
\begin{figure}[ht]
\begin{subfigure}{.9\columnwidth}
    \centering
    \label{fig:model:a}
    \begin{tikzpicture}[scale=2.0]
    \coordinate (C1) at (0,0);
    \coordinate (C2) at (0,-1);
    \coordinate (C3) at (1,0);
    \coordinate (C4) at (0,1);
    \coordinate (C5) at (-1,0);
    \coordinate (C6) at (1,-1);
    \coordinate (C7) at (1,1);
    \coordinate (C8) at (-1,1);
    \coordinate (C9) at (-1,-1);
    \node[graphnode,fill=red] (N1) at (C1) {1};
    \foreach \m in {2,...,9}
    	\node[graphnode] (N\m) at (C\m) {\m};
    \draw (N1) --node[right]{$t$} (N2);
    \draw (N1) --node[above]{$t$} (N3);
    \draw (N1) --node[right]{$t$} (N4);
    \draw (N1) --node[above]{$t$} (N5);
    \draw (N4) --node[above]{$t$} (N7) --node[right]{$t$} (N3) --node[right]{$t$} (N6) --node[above]{$t$} (N2) --node[above]{$t$} (N9) --node[right]{$t$} (N5) --node[right]{$t$} (N8) --node[above]{$t$} (N4);
    \node[above left=5pt,red] at (C1) {$U$};
    \node[below left=5pt] at (C1) {$v_a$};
    \foreach \m in {2,...,5}
        \node[below left=5pt] at (C\m) {$v_b$};
    \foreach \m in {6,...,9}
        \node[below left=5pt] at (C\m) {$v_c$};
    \end{tikzpicture}
\end{subfigure}\\[.5cm]
\begin{subfigure}{.9\columnwidth}
    \centering
    \label{fig:model:b}
    \begin{tikzpicture}[scale=2.0]
    \draw (-1.6,1.6)--(1.6,1.6);
    \coordinate (C1) at (-1,1);
    \coordinate (C2) at (0,1);
    \coordinate (C3) at (1,1);
    \coordinate (C4) at (-1,0);
    \coordinate (C5) at (-1,-1);
    \coordinate (C6) at (0,0);
    \coordinate (C7) at (0,-1);
    \coordinate (C8) at (1,0);
    \coordinate (C9) at (1,-1);
    \node[graphnode,fill=red] (N1) at (C1) {1};
    \foreach \m in {2,...,9}
    	\node[graphnode] (N\m) at (C\m) {\m};
    \draw (N1) --node[above]{$2t$} (N2) --node[above]{$2t$} (N3);
    \draw (N4) --node[right]{$\sqrt{2}t$} (N5);
    \draw (N6) --node[right]{$\sqrt{2}t$} (N7);
    \node[above left=5pt,red] at (C1) {$U$};
    \node[below left=5pt] at (C1) {$v_a$};
    \node[below left=5pt] at (C2) {$v_b$};
    \node[below left=5pt] at (C3) {$v_c$};
    \node[below left=5pt] at (C4) {$v_b$};
    \node[below left=5pt] at (C5) {$v_c$};
    \node[below left=5pt] at (C6) {$v_b$};
    \node[below left=5pt] at (C7) {$v_c$};
    \node[below left=5pt] at (C8) {$v_b$};
    \node[below left=5pt] at (C9) {$v_c$};
    \node[below=10pt, color=blue!50] at (C1) {A};
    \node[above=10pt, color=red!50] at (C4) {B1};
    \node[above=10pt, color=red!50] at (C6) {B2};
    \node[above=10pt, color=yellow!70] at (C8) {C1};
    \node[above=10pt, color=yellow!70] at (C9) {C2};
    \begin{scope}[on background layer]
        \node[fill=blue!15,rounded corners,fit=(N1)(N2)(N3),label=above:] {};
        \node[fill=red!15,rounded corners,fit=(N4)(N5)] {};
        \node[fill=red!15,rounded corners,fit=(N6)(N7)] {};
        \node[fill=yellow!30,rounded corners,fit=(N8)] {};
        \node[fill=yellow!30,rounded corners,fit=(N9)] {};
    \end{scope}
  \end{tikzpicture}
\end{subfigure}
\caption{The Anderson impurity model under consideration. Top panel shows the original lattice, bottom panel the unitarily transformed model where the lattice separates into five independent blocks.}
\label{fig:model}
\end{figure}

Their example is the realization of an Anderson impurity model~\cite{Anderson1961}, depicted in the top panel of Fig.~\ref{fig:model}, with $M=9, N=8$ and the Hamiltonian 
\begin{equation}
    \hat H = t \sum_{i\sim j,\alpha} \aopdag_{i\alpha} \aop_{j\alpha} + \sum_{i\alpha} v_i \aopdag_{i\alpha} \aop_{i\alpha} + U \aopdag_{1\uparrow} \aop_{1\uparrow}\aopdag_{1\downarrow} \aop_{1\downarrow}.
\end{equation}
The first two terms, without the interaction, form the one-particle Hamiltonian $\hat h$.
One always sets $v_{i\up}=v_{i\down}=v_i$.
Further, the potential is assumed to obey $v_1=v_a$, $v_2=v_3=v_4=v_5=v_b$ and $v_6=v_7=v_8=v_9=v_c$ which introduces a $\mathsf{C}_{4\mathsf{v}}$ symmetry to the system. Note that in the case of degeneracy, a ground-state density of this system does not have to necessarily share this symmetry, just the full degeneracy region does. This also means that the density at the middle point of the degeneracy region, the \emph{equidensity} that arises from an equally weighted statistical mixture of the orthonormal basis states of the degenerate subspace (thus leading to fractional occupation numbers), shares the symmetry. The same density is achieved in thermally-assisted-occupation DFT~\cite{Chai2012-TAO-DFT}, where a thermal ensemble is chosen instead of the ground state, if the temperature approximates zero.

The benefit of such a symmetrical setting is that a fully $\mathsf{C}_{4\mathsf{v}}$ symmetric (also taking spin-symmetry into account) density, like the external potential, has only three different components $(\rho_a,\rho_b,\rho_c)$ which allows to plot them in 3D. With the further normalization condition
\begin{equation}\label{eq:condition-density-symmetric}
    \rho_a + 4\rho_b + 4\rho_c = N/2
\end{equation}
such a density is even restricted to a 2D subplane.
Further, a unitary transform $\hat U$ of the one-particle Hamiltonian $\hat h$ in position basis allows to reduce it into a form with five separate blocks, corresponding to the irreducible representations of the point group $\mathsf{C}_{4\mathsf{v}}$. The model and its unitary transformation are displayed in Fig.~\ref{fig:model}, while the unitary matrix that was used can be found in Appendix~\ref{app:unitary}.

Ignoring the Hubbard-$U$ interaction for a moment, this yields a new spin-free one-particle Hamiltonian $\hat h' = \hat U \hat h \hat U^\dagger$ that equally acts on each spin-component.
\begin{equation}\label{eq:block-matrix}
    \hat h' = \begin{pNiceMatrix}
    \Block[fill=blue!15,rounded-corners]{3-3}{}
    v_a & 2t & 0 & 0 & 0 & 0 & 0 & 0 & 0 \\
    2t & v_b & 2t & 0 & 0 & 0 & 0 & 0 & 0 \\
    0 & 2t & v_c & 0 & 0 & 0 & 0 & 0 & 0 \\
    0 & 0 & 0 & \Block[fill=red!15,rounded-corners]{2-2}{} v_b & \sqrt{2}t & 0 & 0 & 0 & 0 \\
    0 & 0 & 0 & \sqrt{2}t & v_c & 0 & 0 & 0 & 0 \\
    0 & 0 & 0 & 0 & 0 & \Block[fill=red!15,rounded-corners]{2-2}{} v_b & \sqrt{2}t & 0 & 0 \\
    0 & 0 & 0 & 0 & 0 & \sqrt{2}t & v_c & 0 & 0 \\
    0 & 0 & 0 & 0 & 0 & 0 & 0 & \Block[fill=yellow!30,rounded-corners]{1-1}{} v_b & 0 \\
    0 & 0 & 0 & 0 & 0 & 0 & 0 & 0 & \Block[fill=yellow!30,rounded-corners]{1-1}{} v_c
    \end{pNiceMatrix}.
\end{equation}
Note not only the difference in the basis ordering compared to \citet{RoesslerVerdozzi2018}, but also the hopping $\sqrt{2}t$ in the $2\times 2$-blocks (red blocks in Eq.~\eqref{eq:block-matrix}, B clusters) that was wrongly put $2t$ in the reference. We will go on with this choice, assuming that it does not influence the effect that was diagnosed as ``non-$v$-representability''. (We also checked the ground state for the hopping $2t$ but there occurred no qualitative difference.)
The important consequence of the transformation is that the Hamiltonian now blocks into five submatrixes that correspond to smaller clusters that do not interact. On the other hand, the on-site energies in the diagonal stay exactly the same while the hopping inside each cluster just gets renormalized. Finally, the interaction from $U$ stays exactly the same since the first lattice site, the only place where it acts between particles of opposing spin, is not changed. Only the first cluster (A cluster, marked in violet) is thus interacting and one can solve for the ground state in each cluster separately, as soon as the respective particle filling for each one is determined, which amounts to a huge reduction in computational complexity.

Now \citet{RoesslerVerdozzi2018} chose the potentials $v_a = -2.69, v_b = -1$ and $v_c=0$ in order to deliberately create a degenerate ground state for the interacting system with $U=1, t=1$ (yet this is not critical for the following example). With these values and a total of $N=8$ particles, one finds the ground states realized with the filling
\begin{equation}
    N_{\rm A}=3, N_{\rm B1}=2, N_{\rm B2}=2, N_{\rm C1}=1, N_{\rm C2}=0.
\end{equation}
Here, particles remain unpaired and thus a spin degeneracy occurs, but it is only `internal' and thus every ground state has the same spin-unresolved density that defines our target density $\rhotarget$.

\begin{figure}[ht]
    \centering
    \includegraphics[width=\columnwidth]{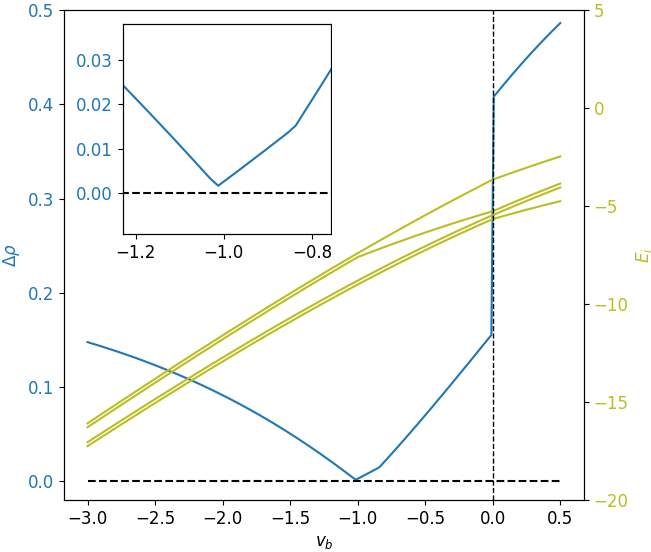}
    \caption{Difference between the target density and the equidensity of each degeneracy region for $v_a=2v_b$, $v_c=0$. A minimum is attained close to $v_b=-1$ but zero is not reached as can be seen in the zoomed-in inset.
    In olive the eigenvalues (slightly displaced for better visibility) of the spin-independent Hamiltonian are plotted to demonstrate the occurrence of double (for $v_b<0$) and triple (at $v_b=0$) ground-state degeneracy.}
    \label{fig:target-dens-distance}
\end{figure}

\begin{figure}[ht]
    \centering
    \includegraphics[width=\columnwidth]{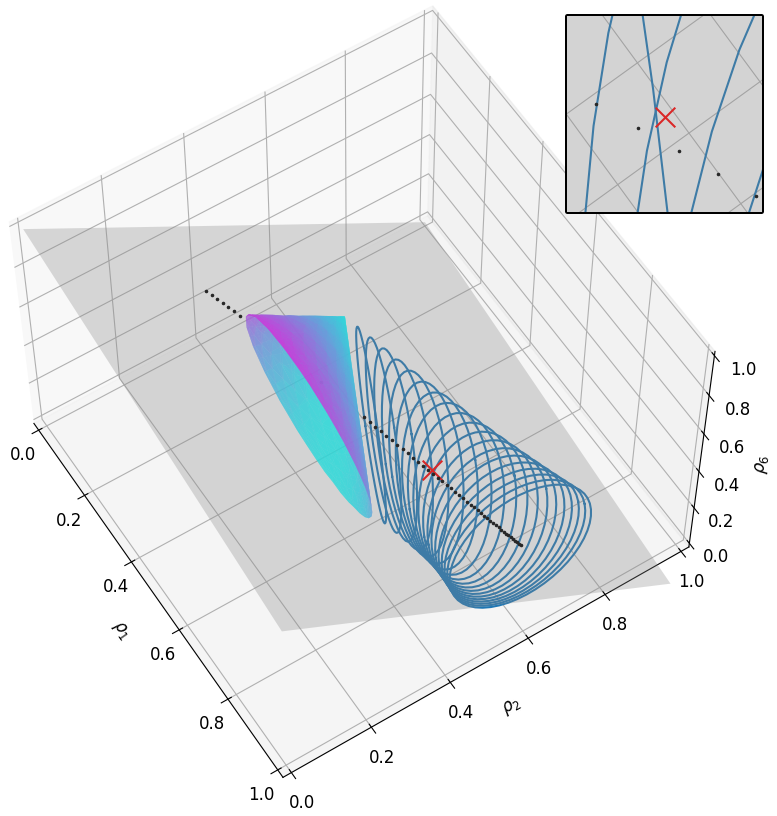}
    \caption{The target density $\rhotarget$ (red cross) is closely missed by the equidensities (black dots; see inset) of the 2D degeneracy regions (blue circles) that form a degeneracy bundle that attaches to the 3D degeneracy region in the shape of a cone. The shaded plane is given by the condition Eq.~\eqref{eq:condition-density-symmetric} for symmetric densities.}
    \label{fig:degen-structure}
\end{figure}

\begin{figure}[ht]
    \centering
    \includegraphics[width=\columnwidth]{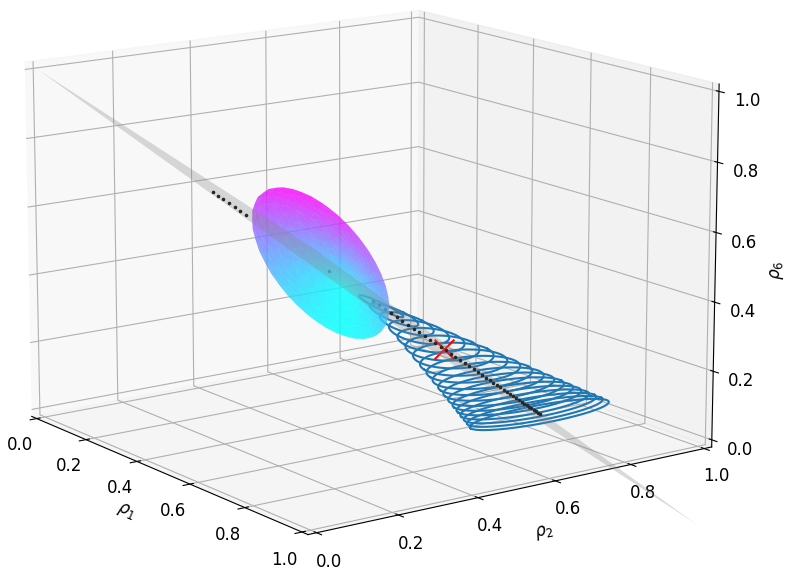}
    \caption{Same as Fig.~\ref{fig:degen-structure} with an in-plane view.}
    \label{fig:degen-structure-2}
\end{figure}

Next, the interaction was turned off ($U=0$) and the potential parameter space was scanned in order to find a representing potential for $\rhotarget$. It is found that with $v_a=2v_b, v_c=0$ degeneracy occurs, but this time the degeneracy regions extend in space and form a degeneracy bundle in the shape of a deformed cylinder. At $v_a=v_b=v_c=0$ the degeneracy region is even 3-dimensional and forms a cone. \citet{RoesslerVerdozzi2018} report that they cannot reach the target density $\rhotarget$ with any of these potentials, and indeed it is not reached if only the equidensity of each degeneracy region is considered, which seems to have been their choice. The difference between $\rhotarget$ and the equidensities is displayed in Fig.~\ref{fig:target-dens-distance} which shows great similarity with Fig.~2 from the reference. The further displayed energy eigenvalues demonstrate how at $v_b=0$ three-fold degeneracy occurs together with a jump of the density difference, since the computed density crosses the whole degeneracy region. But $\rhotarget$ is easily reached with another density from the appropriate degeneracy region since it lies inside the degeneracy bundle, so there is clearly no problem with $v$-representability. The whole situation is visualized in Figs.~\ref{fig:degen-structure} and \ref{fig:degen-structure-2}. All displayed densities are with respect to one spin channel only.

We want to study the appearance of the cone degeneracy region in more detail.
For $U=0$ and maximal symmetry at $v_a=v_b=v_c=0$ a three-fold degeneracy is attained for the ground state in each spin channel. We call the six lowest eigenstates of the (spin-independent) one-particle Hamiltonian Eq.~\eqref{eq:block-matrix} $\phi_1,\ldots,\phi_6$ where $\phi_4,\phi_5,\phi_6$ all have the same eigenvalue. Then with $N/2=4$ particles in each spin channel the degenerate ground-state subspace is spanned by the Slater-determinant states $\Phi_1=\phi_1\wedge\phi_2\wedge\phi_3\wedge\phi_4$, $\Phi_2=\phi_1\wedge\phi_2\wedge\phi_3\wedge\phi_5$, $\Phi_3=\phi_1\wedge\phi_2\wedge\phi_3\wedge\phi_6$ with densities $\orho{1},\orho{2},\orho{3}$. We further define $\oirho{kl}{i} = 2\langle \Phi_k,\hat\rho_i\Phi_l\rangle$ ($k,l\in\{1,2,3\}$, $k\neq l$) which in this special case is zero for all but one choice of indices (say $k=1$, $l=2$) where the result is also real-valued. The degeneracy region is then formed as the convex hull of the set~\cite{penz2023geometry}
\begin{equation}\label{eq:cone-set}
    \left\{ x_1^2\orho{1}+x_2^2\orho{2}+x_3^2\orho{3}+x_1x_2\orho{12} \;\middle|\; \xx\in\R^3, |\xx|=1\right\}.
\end{equation}
Putting in spherical coordinates, $x_1=\sin\theta\cos\varphi$, $x_2=\sin\theta\sin\varphi$, $x_3=\cos\theta$, we can transform the points of the set into
\begin{equation}\begin{aligned}
    &(\sin\theta)^2\left(\frac{\orho{1}+\orho{2}}{2} + \frac{\orho{1}-\orho{2}}{2} \cos(2\varphi) + \frac{\orho{12}}{2}\sin(2\varphi)\right) \\
    +&(\cos\theta)^2\orho{3}.
\end{aligned}\end{equation}
The large bracket in the first line then describes an ellipse with center $(\orho{1}+\orho{2})/2$ that gets double covered because of the appearance of $2\varphi$. Since $(\sin\theta)^2+(\cos\theta)^2=1$ we further have convex combinations between all points of the ellipse and $\orho{3}$ that thus forms the tip of the cone. Equation~\eqref{eq:cone-set} can be seen as the projection of the Veronese variety into projective 3-space (called a Steiner surface), a topic studied by \citet{degen1994} with the cone appearing as (Ae) in his typology. Another famous shape, the Roman surface, was already discovered in previous examples~\cite{penz2023geometry}.


The issue faced by \citet{RoesslerVerdozzi2018} was thus that $\rhotarget$ lies \emph{inside} a degeneracy region of the non-interacting system. The same effect can manifest itself as a convergence issue for the Kohn--Sham method~\cite{KS1965} if the interacting and the auxiliary non-interacting systems have different degeneracy structures. Just like in the example above, the target density might lie within a degeneracy region $D(v_\mathrm{s})$ when the correct exchange-correlation potential is applied to the Kohn--Sham system. This then leads to redundancy when choosing the density.
But already during the iterative approach towards this density, when choosing a potential $v\approx v_\mathrm{s}$, one will always remain \emph{outside} of $D(v_\mathrm{s})$ and just approach the boundary of the degeneracy region as $v\to v_\mathrm{s}$ while never entering it.
Densities inside $D(v_\mathrm{s})$ are thus unattainable by an approximative algorithm if it is not specifically adjusted such that it can enter the degeneracy region. The simplest modification is \emph{damping}~\cite{CANCES_MMNA34_749,cances2001}, where an iteration step, e.g., between densities $\rho_i \to \rho_{i+1}$ (or, analogously, between density matrices) is instead taken as the convex combination
\begin{equation}\label{eq:damping-step}
    \rho_i \to (1-\tau)\rho_i + \tau\rho_{i+1}.
\end{equation}
The step size is chosen such that the energy surely decreases with the finally decided iteration step. Apart from leading to improved convergence of the algorithm (only with such a modification it was possible to show that the Kohn--Sham method in finite dimensions and with additional regularization does in principle converge~\cite{penz2019guaranteed,penz2020erratum}) the step Eq.~\eqref{eq:damping-step} allows to enter inside a degeneracy region also in cases where $\rho_{i+1}$ is always chosen on the outside.

\begin{figure}[ht]
\begin{subfigure}{.9\columnwidth}
    \centering
    \label{fig:random:a}
    \begin{tikzpicture}
        \node[inner sep=0pt] at (0,0)
            {\includegraphics[width=1\columnwidth]{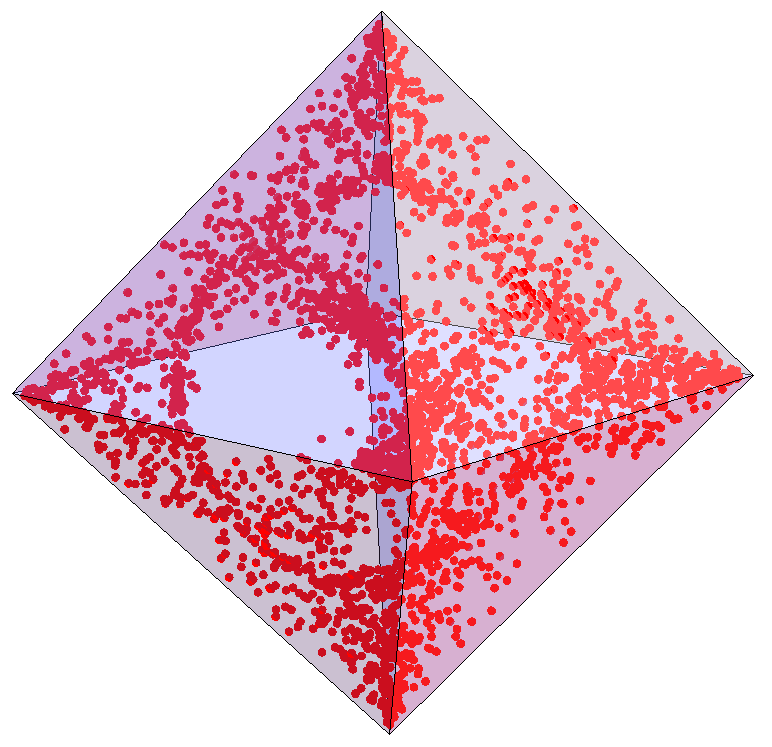}};
        \def\CX{3};
        \def\CY{2.5};
        \def\L{0.8};
        \coordinate (C1) at (0+\CX,{2/3*\L+\CY});
        \coordinate (C2) at ({-2/sqrt(3)*\L+\CX},0+\CY);
        \coordinate (C3) at (0+\CX,{2*\L+\CY});
        \coordinate (C4) at ({2/sqrt(3)*\L+\CX},0+\CY);
        \foreach \m in {1,...,4}
            \node[graphnode] (N\m) at (C\m) {\m};
        \draw (N1) -- (N2);
        \draw (N1) -- (N3);
        \draw (N1) -- (N4);
        \draw (N2) -- (N3);
        \draw (N3) -- (N4);
        \draw (N4) -- (N2);
    \end{tikzpicture}
\end{subfigure}\\[.5cm]
\begin{subfigure}{.9\columnwidth}
    \centering
    \label{fig:random:b}
    \includegraphics[width=1\columnwidth]{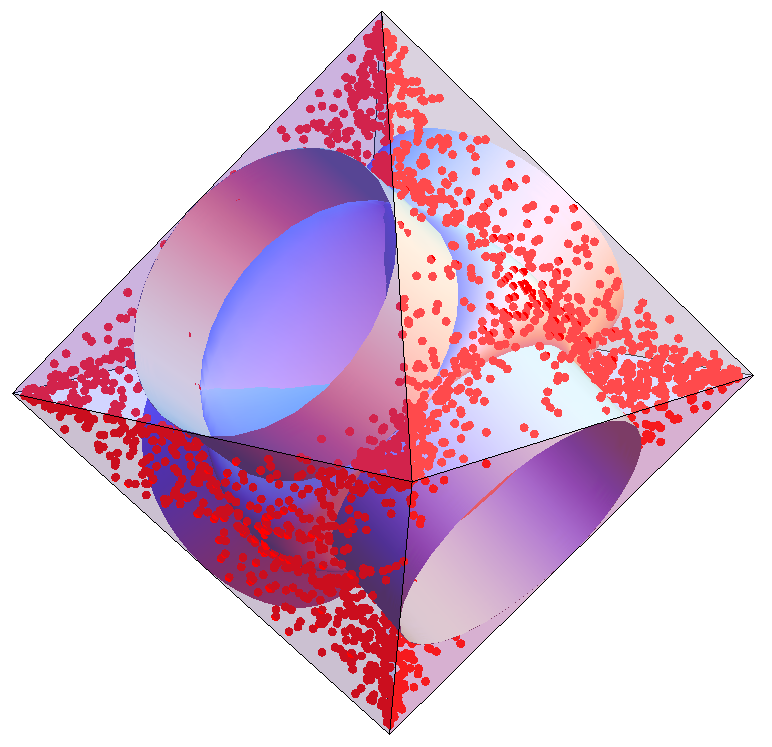}
\end{subfigure}
\caption{Densities of randomly chosen potentials in a $M=4, N=2$ system with tetrahedral symmetry. The lower panel also shows the degeneracy regions.}
\label{fig:random}
\end{figure}
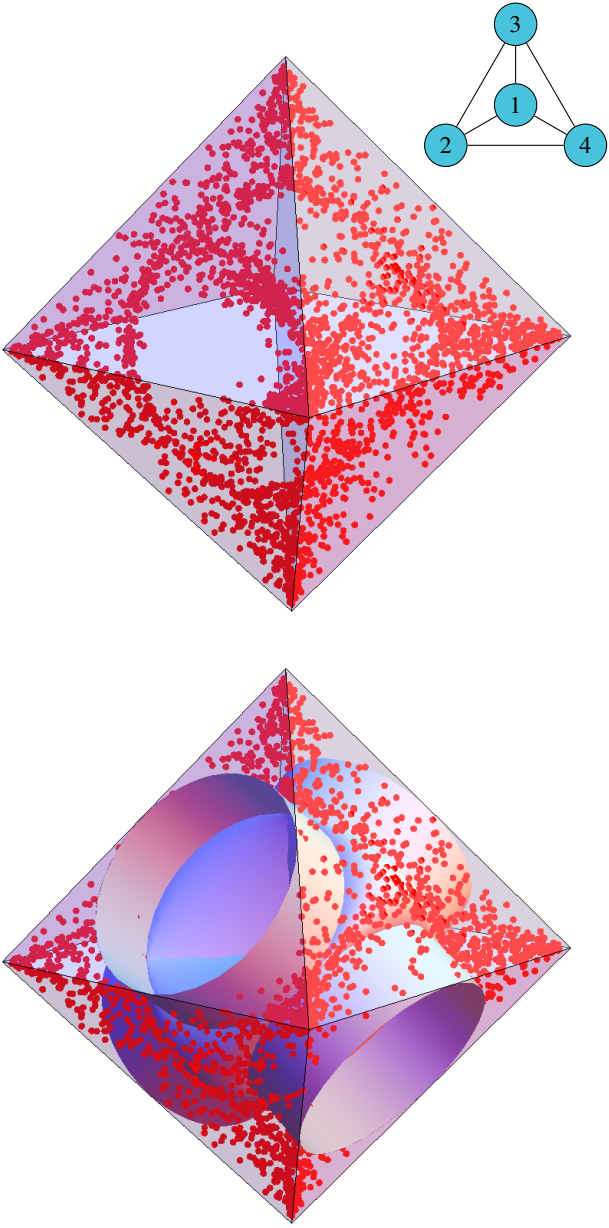

How degeneracy regions would be avoided by a usual Kohn--Sham iteration can easily be visualized by just sampling potentials randomly and plotting their respective densities. Those are almost always unique since the potentials $\PotDeg$ have measure zero in $\Pot$. Figure~\ref{fig:random} shows this situation with an example from Ref.~\onlinecite{penz2023geometry} and how the intricate shape of the degeneracy regions fits into the empty area.

\section{Adiabatic variation of the potential and geometric phase}
\label{sec:adiabatic}


We will now allow for potentials $v(t)$ and $\BB(t)$ that vary in time and for all $t\in[0,T]$ lead to a non-degenerate ground state.
The adiabatic theorem~\cite{sakurai-book} then states that if the system initially is in its ground state and if further this variation is slow enough, it will also remain in the instantaneous ground state of the Hamiltonian $\hat H(t)$. We determine this ground state for the example of a graph with $M=4$ vertices where vertex 1 is linked to all others but no other edges are present. Due to its shape this is called the `claw graph' and it clearly has a trigonal symmetry that can lead to two-fold degeneracy.

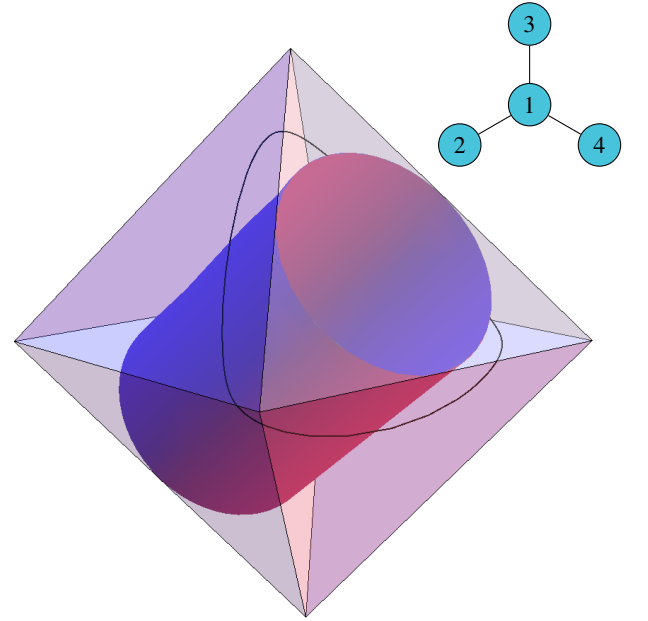
\begin{figure}[ht]
    \centering
    \begin{tikzpicture}
        \node[inner sep=0pt] at (0,0)
            {\includegraphics[width=.9\columnwidth]{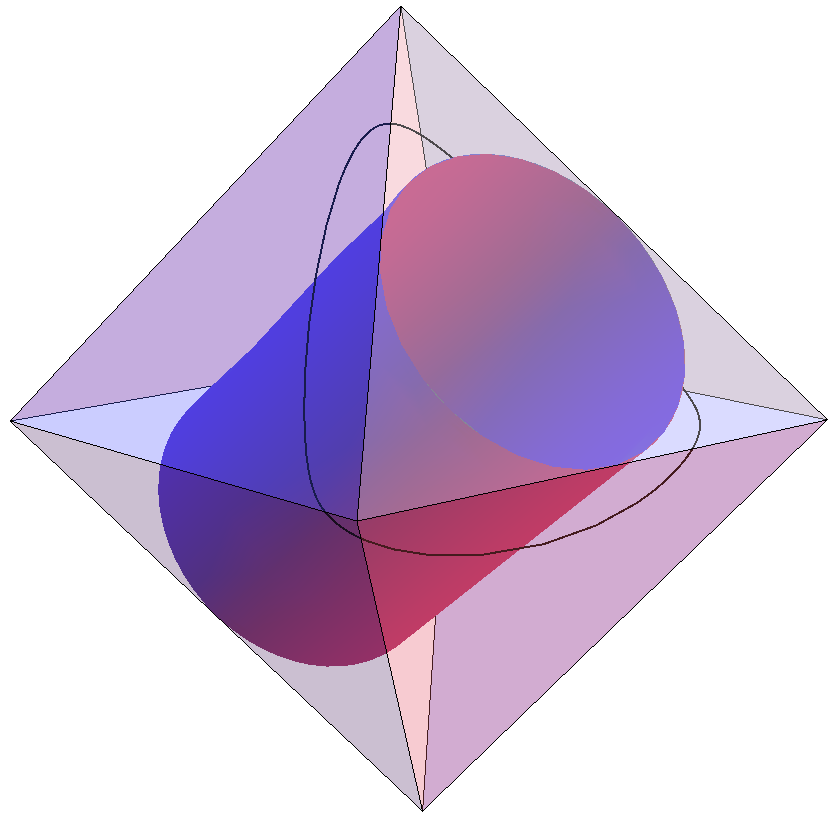}};
        \def\CX{3};
        \def\CY{2.5};
        \def\L{0.8};
        \coordinate (C1) at (0+\CX,{2/3*\L+\CY});
        \coordinate (C2) at ({-2/sqrt(3)*\L+\CX},0+\CY);
        \coordinate (C3) at (0+\CX,{2*\L+\CY});
        \coordinate (C4) at ({2/sqrt(3)*\L+\CX},0+\CY);
        \foreach \m in {1,...,4}
            \node[graphnode] (N\m) at (C\m) {\m};
        \draw (N1) -- (N2);
        \draw (N1) -- (N3);
        \draw (N1) -- (N4);
    \end{tikzpicture}
    \caption{Degeneracy regions (collected into a single degeneracy bundle) of the claw graph with a closed density path that winds them.}
    \label{fig:Oct-path-around-deg}
\end{figure}

If we put $N=2$ particles on the graph then for each spin component this leads to the following two-particle Hamiltonian (following Ref.~\onlinecite{penz2021DFTgraphs}),
\begin{equation}
\hat H_0=
\begin{pmatrix*}[r]
4 & 0 & 0 & 1 & 1 & 0 \\
0 & 4 & 0 & -1 & 0 & 1 \\
0 & 0 & 4 & 0 & -1 & -1 \\
1 & -1 & 0 & 2 & 0 & 0 \\
1 & 0 & -1 & 0 & 2 & 0 \\
0 & 1 & -1 & 0 & 0 & 2
\end{pmatrix*}.
\end{equation}
We put $\BB(t)=0$ and the (time-dependent) external potential $v=(v_1,v_2,v_3,v_4)$ adds $(v_1+v_2,v_1+v_3,v_1+v_4,v_2+v_3,v_2+v_4,v_3+v_4)$ along the diagonal of the Hamiltonian to yield $\hat H(t)$. We can easily test that for $v=0$ the ground state is two-fold degenerate and that changing just $v_1$ (that acts on the center vertex 1) does not change the symmetry of the Hamiltonian, thus this degeneracy persists while the ground-state density changes. The degeneracy region for any $v=(s,0,0,0)$, $s\in\R$, is a precise circle (the two-fold degeneracy regions always give ellipses as degeneracy regions \cite[Sec.~III]{penz2023geometry}) and those circles are stacked above each other to form a filled cylinder that runs diagonally through the whole density domain in the shape of a octahedron (see Fig.~\ref{fig:Oct-path-around-deg}). Note that in this case the degeneracy regions come arbitrarily close to each other and to the boundary of the density domain (for $|s|\to\infty$) but never touch each other nor the boundary.

\begin{figure}[ht]
    \centering
    \includegraphics[width=1\columnwidth]{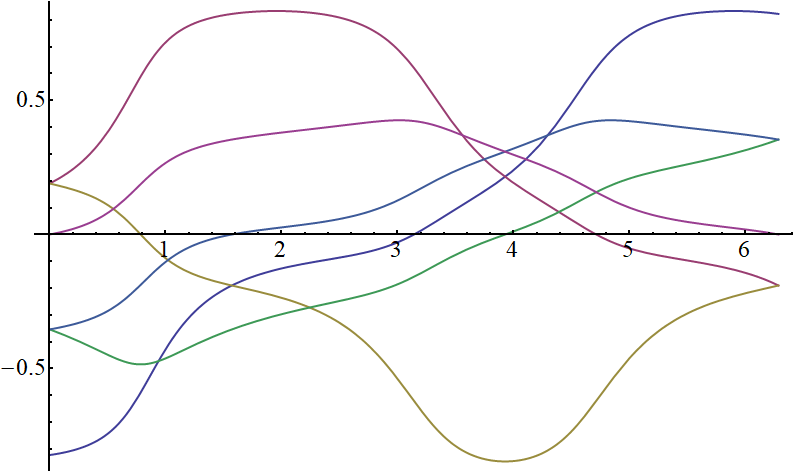}
    \caption{The six components of the instantaneous ground-state wave function to $\hat H(t)$ with a potential winding the degeneracy bundle. Note the sign jump when comparing the values at 0 and $2\pi$.}
    \label{fig:Psi-components-jump}
\end{figure}

We now choose a closed path $v(t)$, $t\in[0,2\pi]$, in the potential space that avoids the degeneracy at $v=(s,0,0,0)$,
\begin{equation}
    v(t) = \cos t \; (0,1,0,0) + \sin t \; (0,0,1,0).
\end{equation}
How this path looks in density space can be seen in Fig.~\ref{fig:Oct-path-around-deg}, where we note that it winds around the cylindrical degeneracy bundle such that it cannot be contracted to a point without cutting through a degeneracy region.
Let $\Psi(t)$ be the instantaneous ground state of $\hat H(t)$ that is unique up to a complex global phase. We will choose it real in all its components (which is always possible since the Hamiltonian is real) and Fig.~\ref{fig:Psi-components-jump} shows its six components.
On first sight these look like nice continuous functions, but if one compares closely the values at $t=0$ and $t=T=2\pi$, we see that a sign jump occurs. This means that even though $\hat H(t)$ was varied continuously and  has $\hat H(0)=\hat H(T)$, the corresponding $\Psi(t)$ is \emph{not} continuous but includes a $\pi$ phase jump. By lifting our condition of a real ground state and including a complex global phase $\rme^{\rmi\varphi/2}$ this phenomenon can be cured, but we still include a phase change of $\pi$ while going from $t=0$ to $t=2\pi$. This phase can be calculated as 
\begin{equation}
    \gamma(T) =  \rmi \int_0^T \langle \Psi (t), \partial_t \Psi (t) \rangle \,\rmd t
\end{equation}
and is exactly the geometric phase introduced by \citet{Berry1984}. The reason for the phase accumulation is that the chosen path in density space cannot be continuously contracted to a point (it is not homotopic to a point) without intersecting a degeneracy region.

Note that if we choose a path that does not have this property, i.e., can be continuously contracted to a point without intersecting a degeneracy region (it is homotopic to a point), then no phase accumulation occurs and we can always choose a real \emph{and} continuous $\Psi(t)$. This offers the possibility to study the occurrence of degeneracy regions/bundles and the fundamental group of $\PotNonDeg$ by calculating the accumulated phase along different closed potential curves.

\section{Time-dependent setting}
\label{sec:td}

After having studied in detail the ground state theory we now turn our attention to the time-dependent version of the theory. 
A particularly interesting question in connection with the degeneracy structure is how a potential that is used to slowly steer the density through a degeneracy region has to look like. We address this question at the end of this section and after first giving a general discussion of time-dependent density-functional theory (TDDFT)~\cite{ullrich-book,Ullrich2013-TDDFT}, but restricted to the lattice setting~\cite{li2008time,farzanehpour2012time}.

Therein, we aim at solving the time-dependent Schr\"odinger equation 
\begin{equation}
    \rmi \partial_t \Psi  (t) = \hat{H} (t) \Psi (t)
    \label{TDSE}
\end{equation}
where the Hamiltonian is any one of the forms Eqs.~\eqref{eq:ham-4component}-\eqref{eq:ham-usual-DFT} and in general we allow for time-dependent potentials. However, for simplicity, let us restrict to the spinless case in the following and consider a general time-evolution with potential $v (t)=(v_1 (t), \ldots,v_M (t))$ starting from a given initial state $\Psi_0$ at time $t_0$.
Suppose we are given a time-dependent density $\rho (t)$
and an initial state $\Psi_0$ with density $\rho(t_0)$. Then we can ask whether there exists
a potential $v(t)$ that generates $\rho (t)$ from solving the time-dependent Schr\"odinger equation with initial state $\Psi_0$ and using $v(t)$ as potential. This is the time-dependent $v$-representability problem.
In the continuum case, this problem has been widely studied~\cite{Runge1984,vanLeeuwen1999,Ruggenthaler_2011,ullrich-book} and most proofs rely on temporal analyticity~\cite{Fournais2016} although mathematically rigorous proofs are still lacking, see Ref.~\onlinecite{Ruggenthaler_2015} for a review.
In the lattice case the situation is more favorable and some precise statements related to $v$-representability can be made. We will here outline the main results.
We take a time-dependent Hamiltonian of the form
\begin{equation}
    \hat H (t) = \sum_{i,j} h_{ij} \aopdag_{i} \aop_{j} + \frac{1}{2}\sum_{i,j} w_{ij} \aopdag_{i} \aopdag_{j}\aop_{j} \aop_{i}  + \sum_{i} v_i  (t) \aopdag_{i} \aop_{i}.
\end{equation}
with $w_{ij}=w_{ji}$ real and symmetric.
Applying the Heisenberg equation of motion to
the density operator and denoting
$\rho_k (t)=\langle \hat{\rho}_k \rangle_{\Psi(t)}$ we
get the lattice-equivalent of the continuity equation,
\begin{equation}\label{continuity}
\begin{aligned}
\partial_t \rho_k &= -\rmi \langle [\hat{\rho}_k, \hat{H}(t)] \rangle_{\Psi(t)} \\
&= \sum_j 2\Im h_{kj} \gamma_{jk} (t) 
= - \sum_j J_{kj} (t),
\end{aligned}
\end{equation}
where we defined the one-particle density matrix
$\gamma_{jk}(t) = \langle \aopdag_k \aop_j \rangle_{\Psi(t)}$
and the bond current
$J_{kj}(t) = 2 \Im h_{jk} \gamma_{kj} (t)$.
Using the Heisenberg equation of motion a second time and working out the commutators gives
\begin{equation}\label{v_eq1}
\begin{aligned}
    \partial_t^2 \rho_k (t) &= -\rmi \partial_t
    \langle [\hat{\rho}_k, \hat{H}(t)] \rangle_{\Psi(t)} \\
    &= -
    \langle [[\hat{\rho}_k,\hat{H}(t)], \hat{H}(t)] \rangle_{\Psi(t)} \\
    &= q_k (t) + \sum_{j} K_{kj} (t) v_j (t),
\end{aligned}
\end{equation}
where we defined
\begin{equation}
    q_k (t) =  -\sum_j 2\Re h_{kj}
    \langle [\aopdag_k \aop_j, \hat{H}_0]\rangle_{\Psi(t)}
\end{equation}
and
\begin{equation}
    K_{kj} (t) = 2\Re\left( -h_{jk} \gamma_{kj} (t) + 
    \delta_{kj} \sum_i h_{ji} \gamma_{ij} (t) \right)
    \label{K_mat}
\end{equation}
If we choose the gauge $v_M(t)=0$, we can rewrite
Eq.~\eqref{v_eq1} as
\begin{equation}
    \sum_{j=1}^{M-1} K_{kj} (t) v_j (t) = \partial_t^2 \rho_k (t) - q_k (t).
    \label{v_eq}
\end{equation}
If we are given a density $\rho_k (t)$ over time we can prescribe $\partial_t^2 \rho_k$ on the right-hand side of this equation and try solve for $v$ at any $t> t_0$ when the matrix $K$ and the vector $q$ are known. The latter quantities are calculable from the wave function, which in turn dependents on $v$ at all previous times. It was shown by \citet{farzanehpour2012time} that solving Eq.~\eqref{v_eq} together with the time-dependent Schrödinger equation allows for a $v$-representability solution for a sufficiently short time interval $[t_0,t_0+\Delta t]$, provided that $K(t_0)$ is invertible. This procedure then yields a potential that produces the prescribed density in this time interval.
The natural question to ask is how the size of $\Delta t$ is determined. It is known that on lattices the density becomes non-$v$-representable when it changes too fast 
from one lattice site to the next~\cite{li2008time,farzanehpour2012time}. To see this, let us assume that $\partial_t \rho_k$ in Eq.~\eqref{continuity} becomes large.
From the definition of the bond current we see that its maximal
value is reached when the imaginary parts of $h_{jk} \gamma_{kj} $ for different $j$ are maximal, i.e., when $\Re h_{jk} \gamma_{kj} =0$ for all $j$.
But this happens precisely when $K_{kj}=K_{jk}=0$ for all $j \neq k$ in which case it follows from Eq.~\eqref{K_mat} that also $K_{kk}=0$; thereby the $k$th row and column of the matrix
vanishes and $K$ becomes non-invertible~\cite{farzanehpour2012time}.
To specify a $v$-representability domain we therefore have to restrict the density
changes on neighboring sites. 
One way would be to prove that under a condition of the type 
\begin{equation}
    \left|\frac{\partial_t \rho_i}{\rho_i} \cdot \frac{\partial_t \rho_j}{\rho_j}\right| \leq C |h_{ij}|    
\end{equation}
for some constant $C$ the matrix $K$ is invertible at all times. In such a case $v$-representability can continue up to indefinite times.
Such a condition was explicitly derived for a simple dimer system~\cite{farzanehpour2012time}, however for a general lattice this remains an unproven conjecture.

We will next consider the linear-response regime. Let $v(t)$
and $v_\s (t)$ be the potentials of an interacting and a non-interacting system, respectively, that share the same density $\rho(t)$ and that both evolve from initial states with the appropriate density. Also in the time-dependent setting, the non-interacting system is called the Kohn--Sham system~\cite{ullrich-book,Ullrich2013-TDDFT} and its external potential is typically split up as 
$v_\s (t)=v(t)+ v_\Hxc (t)$ which defines the Hartree-exchange-correlation potential $v_\Hxc (t)$. Functional differentiation of $v_\s (t)$ with respect to the density (assuming that this can be appropriately defined) yields
\begin{equation}
    \frac{\delta v_{\s,i} (t)}{\delta \rho_j (t')} = \frac{\delta v_{i} (t)}{\delta \rho_j (t')} + \frac{\delta v_{\Hxc,i} (t)}{\delta \rho_j (t')}.
    \label{response1}
\end{equation}
If we define the functions
\begin{align}
    \chi_{ij} (t,t')&= \frac{\delta \rho_{i} (t)}{\delta v_j (t')} \\
     \chi_{\rms,ij} (t,t')& = \frac{\delta \rho_{i} (t)}{\delta v_{\rms,j} (t')} \\ 
     f_{\Hxc,ij} (t,t') &= \frac{\delta v_{\Hxc,i} (t)}{\delta \rho_j (t')},
\end{align}
where $\chi$ and $\chi_\s$ are the density response functions of the interacting and the Kohn--Sham system and $f_\Hxc$ is the Hartree-exchange-correlation kernel. We can rewrite Eq.~\eqref{response1} in the form
\begin{equation}
    \chi = \chi_\s + \chi_\s \cdot f_\Hxc \cdot \chi,
    \label{response2}
\end{equation}
where the product is defined as a matrix product in the lattice indices and with an integration over time variables.  Eq.~\eqref{response2} is commonly used to calculate the excitation energies of the interacting system, which appear as poles of $\chi$ after a Fourier transformation from time to frequency domain. Similarly $\chi_\s$ contains the excitation energies of the Kohn--Sham system. When an approximation for $f_\Hxc$ is given, the excitation energies of the interacting system can then be obtained from a Kohn--Sham calculation.

We now want to make a connection between this formalism and the discussion of degeneracy. When we perturb the ground state of a non-degenerate system with a small time-dependent $\delta v(t)$, the response function $\chi$ is well-defined. However, if the ground state is degenerate, poles in the response function merge and make it ill-defined. If we assume that the changes are adiabatic, this can be related to Rellich's theorem~\cite{penz2021DFTgraphs,rellich-book,rellich1937}: when a degenerate ground state is perturbed dependent on a small prefactor $\lambda$, then Rellich's theorem asserts that there exists a possible choice within the set of degenerate states that depends analytically on $\lambda$, although most choices will not. Physically this means that the response is only defined for changes in very special directions in potential space. In practice, however, almost all applications of linear-response TDDFT are carried out for systems 
with non-degenerate ground states. For the calculation of excitation energies of systems with degenerate ground states an alternative approach based on choosing a non-degenerate excited state as a reference has been suggested instead~\cite{Seth2005}.

\begin{figure}[ht]
    \centering
    \begin{tikzpicture}[scale=0.8]
        \draw[->] (0, 0) -- (6.5, 0) node[below] {};
        \draw[->] (0, 0) -- (0, 4) node[left] {$E$};
        \draw[-, MidnightBlue, thick] (0, 3.5) -- (6, 0.5);
        \draw[-, MidnightBlue, thick] (0, 0.5) -- (6, 3.5);
        \draw[-,dashed] (1, 0) -- (1, 6.5);
        \draw[-,dashed] (5, 0) -- (5, 6.5);
        \draw[decoration={brace,raise=2pt},decorate] (1,1) -- node[left=4pt] {$\Delta E$} (1,3);
        \draw[decoration={brace,mirror,raise=2pt},decorate] (5,1) -- node[right=4pt] {$\Delta E$} (5,3);
        \fill[red] (3, 2) circle (2.5pt);
        \node[below] at (1, 0) {$v_1$};
        \node[below] at (5, 0) {$v_2$};
        \node[above] at (1, 6.5) {$\rho_1$};
        \node[above] at (5, 6.5) {$\rho_2$};
        \draw[-,red,dashed] (3, 2) -- (1.59, 6);
        \draw[-,red,dashed] (3, 2) -- (4.41, 6);
        \fill[red,opacity=0.15] (3, 6.5) circle (1.5);
        \draw[red] (3, 6.5) circle (1.5); 
        \fill[MidnightBlue] (5, 6.52) circle (2.5pt);
        \fill[MidnightBlue] (1, 6.48) circle (2.5pt);
        \draw[-, thick, dashed, MidnightBlue] (1.5, 6.5) to (4.5, 6.5);
        \draw[domain=0:2, smooth, thick, variable=\x, MidnightBlue] plot ({-\x+1.5}, {-0.1*\x*\x+6.5});
        \draw[domain=0:2, smooth, thick, variable=\x, MidnightBlue] plot ({\x+4.5}, {0.1*\x*\x+6.5});
    \end{tikzpicture}
    \caption{A time-dependent density path through a degeneracy region on the top, where densities just outside the degeneracy region correspond to potentials with a (small) energy gap $\Delta E$.}
    \label{fig:crossing-delta-E}
\end{figure}
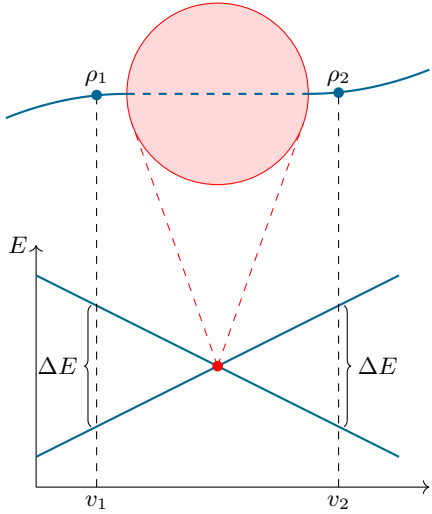

Let us finally come back to the question asked at the beginning of the section. Suppose we consider two densities $\rho_1$  and $\rho_2$, just outside a degeneracy region, and connect them with a smooth path $\rho(t)$ that crosses the degeneracy region. Let us further assume that the path is traversed in a slow manner.
We can then ask for the time-dependent potential $v(t)$ that produces the given density path from a solution of the time-dependent Schrödinger equation. Since the starting and end densities are just outside a degeneracy region, the static potentials $v_1$ and $v_2$ that have $\rho_1$ and $\rho_2$ as their respective ground states correspond to nearly degenerate systems. The whole situation is depicted in Fig.~\ref{fig:crossing-delta-E}. Since the adiabatic theorem can only be applied when the change in the potential is small compared to the energy gap $\Delta E$ between the ground and excited states, we cannot reliably use the adiabatic theorem to derive an approximation for $v(t)$. Such a study of driving the density slowly into a degeneracy region was carried out by \citet{RoesslerVerdozzi2018} and it was found that the potential will exhibit fast oscillations to produce a nearly time-constant density. 
For this reason we can expect that the presence of degeneracy regions has a noticeable effect on the behavior of the potential in TDDFT calculations.
Apart from this pioneering work, we are unaware of studies on this issue.
The implications of the degeneracy structure in density space for TDDFT is still largely unexplored and further research is warranted.

\section{Outlook}
\label{sec:outlook}

By offering a geometrical perspective on fundamental aspects of spin-lattice DFT we hope to set the stage for further inquiries in this direction as well as stimulate contributions from a wider mathematical field. Even though there are clear links to graph theory and algebraic geometry, those areas remain mostly untapped. 
We close with a list of open and unfinished topics that continue the direction of research started here.

\begin{itemize}
    \item We showed how simple spin-lattice systems that are ubiquitous as model systems, like the Anderson impurity model described in Sec.~\ref{sec:verdozzi-model}, can be studied with respect to their degeneracy geometry. Yet, we did not develop tools to efficiently and reliably find all degeneracy regions. As far as degeneracy due to symmetry is concerned, the group-theoretic tools mentioned in Sec.~\ref{sec:setting} and used in Sec.~\ref{sec:verdozzi-model} should do the trick. Additionally, the geometric phase from closed integral contours in potential space provides an indicator for degeneracy regions and relates to the fundamental group of $\PotNonDeg$, as mentioned in Sec.~\ref{sec:adiabatic}.
    
    \item In this and the preceding works on the subject~\cite{penz2021DFTgraphs,penz2023geometry} already several different degeneracy regions up to $\dim D=3$ have been discovered and described. They range from ellipses to cones and the convex hull of the Roman surface. In Ref.~\onlinecite{penz2023geometry} we also suggested a first classification scheme for degeneracy regions in terms of the degree of degeneracy and the nullity of the map from the Veronese variety to density space. This scheme can definitely be refined and many more examples of intricate degeneracy regions are possibly still to be found.
    
    \item The central objects of the Kohn--Sham method in DFT are the exchange-correlation density functional $F_{\rm Hxc}(\rho)$ and its functional derivative $v_{\rm Hxc}(\rho)$. The relative simplicity of the discussed spin-lattice systems allows it to study these objects in-depth, also along the `adiabatic connection'~\cite{Perdew2001}. Degeneracy regions clearly play a role here, since on them $F(\rho)$ has a constant derivative. A first simple yet non-trivial trial system could be the Hubbard trimer at half filling, for which a detailed study about the magnetization density functional is already available~\cite{Ullrich2019-trimer}.
    
    \item Due to the availability of a rigorous HK result in the continuum~\cite{Garrigue2018}, a $v$-representable density is always \emph{uniquely} $v$-representable there. On the other hand, and contrary to the lattice setting, no real criterion for $v$-representability is available within the usual $L^p$-space formulation. A possible remedy is to switch to the Sobolev space $H^1$ for densities, where indeed $v$-representability holds for strictly positive densities on a one-dimensional ring if also distributional potentials are considered~\cite{sutter2023solution}. Yet, this formulation again opens the general possibility of non-unique $v$-representability, just like in the lattice case, but up to now no examples are known. The precise connection between lattices and this setting, in the form of a continuum limit and with respect to degeneracy regions and related concepts, would be a further interesting object of study.
    
    \item There have been considerable advances in the study of quantum-transport phenomena using lattice DFT, as reviewed by \citet{Kurth_2017}. An important application was the description of the Kondo effect in transport through a single Anderson impurity, where it was found that the Kohn--Sham potential becomes a discontinuous function of the site occupation number
    in the zero-temperature limit~\cite{Kurth_2016}.
    For the case of larger lattices, such as the Anderson impurity model discussed in this perspective, the Kohn--Sham potential is likely to reveal structures related to the degeneracy
    regions which poses certain challenges for the construction of future approximate DFT functionals for quantum transport.
    
    \item For the case of time-dependent DFT, the limitations of the currently available $v$-representability proofs have been laid out in Sec.~\ref{sec:td} together with a possible strategy how to overcome them. The next necessary step would then be a convergence proof for the corresponding time-dependent Kohn--Sham method. We further conjectured a noticeable effect due to degeneracy on the (slow) steering of a density through degeneracy regions that has yet to be investigated.
    
    \item A rigorous treatment of finite-temperature lattice DFT was given by \citet{CCR1985} in which it was proven that any density in the interior of the density domain is uniquely $v$-representable.
    In the low temperature limit the density functionals are found to be strongly dependent the lattice occupation number~\cite{Xianlong2012}, which is intimately connected to the filling of
    levels according to the magnitude of the chemical potential. Clearly, here both the degeneracy structure and the particle number play a crucial role. While the dependence on particle number has received considerable attention~\cite{Xianlong2012,Kurth_2017,Sobrino_2020,Sobrino_2022}, the study of degeneracy in this context remains largely unexplored.
    
    \item The states corresponding to the corners of the hypersimplices that form the density domains for pure-lattice systems are naturally encoded as multi-qubit basis states on a quantum computer~\cite{nielsen-book}. For an arbitrary quantum state $\Psi$ the probability that qubit $i$ has value $1$ is given by the density $\rho_i$, such that our geometrical representation may provide a useful setting for quantum-computing applications. 
    Given this natural setting, it is also conceivable that the minimization needed for the constrained-search functional $F(\rho)$ on our lattice systems can be efficiently implemented using a quantum algorithm, as was recently done for the Hubbard dimer~\cite{Pemmaraju_2022}. Since spin-lattices have special interest in the quantum computing community, further connections to the field of quantum information~\cite{Aliverti_2024} seem worthwhile to explore.
\end{itemize}

\section*{Data availability statement}

The plots were created with the help of Python and Mathematica scripts that can be downloaded at {\small\url{https://mage.uber.space/dokuwiki/material/fermion-graph}}. All further data that support the findings of this study are available from the corresponding author upon reasonable request.

\acknowledgments
RvL acknowledges support from the Finnish Academy under project number 356906.

\appendix
\onecolumngrid
\section{Unitary transformation of the Anderson impurity model}
\label{app:unitary}

Here, we give the unitary transformation that leads to a block-diagonalization of the model discussed in Sec.~\ref{sec:verdozzi-model}.
Additionally to what is defined in \citet[Eq.~(2)]{RoesslerVerdozzi2018} it also includes a permutation of the basis vectors and a removal of imaginary off-diagonal contributions in the $2\times 2$-blocks.

\begin{equation}
    \hat U = \frac{1}{2}\begin{pmatrix}
    2 & 0 & 0 & 0 & 0 & 0 & 0 & 0 & 0 \\
    0 & 1 & 1 & 1 & 1 & 0 & 0 & 0 & 0 \\
    0 & 0 & 0 & 0 & 0 & 1 & 1 & 1 & 1 \\
    0 & -1 & -\rmi & 1 & \rmi & 0 & 0 & 0 & 0 \\
    0 & 0 & 0 & 0 & 0 & -(-1)^{1/4} & -(-1)^{3/4} & (-1)^{1/4} & (-1)^{3/4} \\
    0 & -(-1)^{1/4} & (-1)^{3/4} & (-1)^{1/4} & -(-1)^{3/4} & 0 & 0 & 0 & 0 \\
    0 & 0 & 0 & 0 & 0 & -1 & \rmi & 1 & -\rmi \\
    0 & 1 & -1 & 1 & -1 & 0 & 0 & 0 & 0 \\
    0 & 0 & 0 & 0 & 0 & 1 & -1 & 1 & -1 
    \end{pmatrix}.
\end{equation}

\end{document}